\documentclass[letter,11pt]{article}

\pdfoutput=1 

\usepackage{jheppub} 

\usepackage[sort&compress]{natbib} 
\usepackage{amsmath}
\usepackage{graphicx}
\usepackage{dcolumn}
\usepackage{bm}
\usepackage{bbm}
\usepackage{epsfig}
\usepackage{slashed}
\usepackage{amssymb}
\usepackage{color}
\usepackage[font=small]{caption}
\usepackage[font=small]{subcaption}
\usepackage{tabulary}
\usepackage{soul}
\usepackage{multirow}
\usepackage[normalem]{ulem}

\makeatletter

\newcommand{\newc}{\newcommand}
\newc{\renewc}{\renewcommand}

\newcommand{\frules}{{\sc Feyn\-Rules}}

\def\beq{\begin{equation}}
\def\eeq{\end{equation}}
\def\bea{\begin{eqnarray}}
\def\eea{\end{eqnarray}}
\def\bitem{\begin{itemize}}
\def\eitem{\end{itemize}}
\def\ba{\begin{array}}
\def\ea{\end{array}}
\def\bal{\begin{align}}
\def\eal{\end{align}}
\def\bi{\begin{itemize}}
\def\ei{\end{itemize}}
\def\lsim{\mathrel{\rlap{\lower4pt\hbox{\hskip1pt$\sim$}}
    \raise1pt\hbox{$<$}}}         
\def\gsim{\mathrel{\rlap{\lower4pt\hbox{\hskip1pt$\sim$}}
    \raise1pt\hbox{$>$}}}
\newc{\red}{\textcolor{red}}
\newc{\blue}{\textcolor{blue}}

\newc{\ie}{{\it i.e.~}}          \newc{\etal}{{\it et al.~}}
\newc{\eg}{{\it e.g.~}}          \newc{\etc}{{\it etc.~}}
\newc{\cf}{{\it c.f.~}}
\newc{\vs}{{\it vs.~}}
\newc{\os}{\mbox{\hspace{4pt}}}
\newc{\us}{\mbox{\hspace{12pt}}}

\renewc{\bar}{\overline}

\newc{\gev}{\,{\rm GeV}}
\newc{\mev}{\,{\rm MeV}}
\newc{\ev}{\,{\rm eV}}
\newc{\kev}{\,{\rm keV}}
\newc{\tev}{\,{\rm TeV}}
\def\ln{\mathop{\rm ln}}

\newc{\LM}{\mathcal{L}}
\newc{\SM}{\mathcal{S}}

\newc{\HM}{\mathcal{H}}
\newc{\GM}{\mathcal{G}}
\newc{\OM}{\mathcal{O}}
\newc{\FM}{\mathcal{F}}
\newc{\AM}{\mathcal{A}}
\newc{\BM}{\mathcal{B}}
\newc{\NM}{\mathcal{N}}
\newc{\WM}{\mathcal{W}}
\newc{\ZM}{\mathcal{Z}}

\newc{\Chi}{\mathcal{X}}

\definecolor{red1}{cmyk}{0,1,1,0.1}
\definecolor{blue1}{cmyk}{1,0,0,0}

\newcommand{\mpt}{{\;/\!\!\!\! \vec{P}_T}} 

\title{Probing the Top-Higgs Yukawa CP Structure in dileptonic $t \bar t h$ with $M_2$-Assisted  Reconstruction}

\author[a]{Dorival Gon\c{c}alves,}
\author[b]{Jeong Han Kim,}
\author[b]{Kyoungchul Kong}

\affiliation[a]{PITT-PACC,  Department of Physics and Astronomy, University of Pittsburgh, USA}
\affiliation[b]{Department of Physics and Astronomy, University of Kansas, Lawrence, KS 66045, USA}

\preprint{PITT-PACC-1807}

\emailAdd{dorival.goncalves@pitt.edu}
\emailAdd{jeonghan.kim@ku.edu} 
\emailAdd{kckong@ku.edu}

\abstract{Constraining the Higgs boson properties is a cornerstone of the LHC program. We study the potential 
to directly probe the Higgs-top CP-structure via the $t\bar{t}h$ channel  at the LHC with the Higgs boson decaying
to a bottom pair and top-quarks in the dileptonic mode. We show that a combination of laboratory  and $t\bar{t}$ rest frame 
observables display large CP-sensitivity,  exploring the spin correlations in the top decays. To efficiently reconstruct our final
state, we present a method based on simple mass minimization and prove its robustness  to shower, hadronization and 
detector effects. In addition, the mass reconstruction works as an extra relevant handle for background suppression. Based
on our results, we demonstrate that the Higgs-top CP-phase $(\alpha)$  can be probed up to $\cos\alpha< 0.7$ at the high luminosity LHC.
}


\begin{document}

\maketitle
\flushbottom

\section{Introduction}

After the discovery of the Higgs boson at the Large Hadron Collider (LHC)~\cite{Aad:2012tfa,Chatrchyan:2012ufa},
the determination of its properties has become a prominent path in the search for physics beyond the Standard 
Model (SM)~\cite{Higgs:1964ia,Higgs:1964pj,Englert:1964et}. So far, measurements based on the
Higgs signal strengths conform to the SM predictions~\cite{Khachatryan:2016vau,Corbett:2015ksa}. However, the tensor 
structure of the Higgs couplings to other matter fields remains relatively unconstrained. A particularly interesting option 
is that the Higgs interactions present new sources of CP-violation, which could be a key element in explaining the
matter-antimatter unbalance in the Universe~\cite{Sakharov:1967dj,Espinosa:2011eu}.

CP-violation in the Higgs sector has been searched for at the LHC mostly via Higgs couplings with $W^\pm$ and $Z$ gauge 
bosons throughout the Higgs  decays $h\rightarrow W^+W^-$ and $ZZ$~\cite{Plehn:2001nj,Hagiwara:2009wt,Bolognesi:2012mm,
Englert:2012xt,Freitas:2012kw,Ellis:2012wg,Ellis:2012jv,Englert:2013opa,Khachatryan:2014kca,Brehmer:2017lrt}. However, 
these possible CP-violating interactions are one-loop suppressed, arising  only via operators of dimension-6 or
higher~\cite{Buchmuller:1985jz,Grzadkowski:2010es}. On the other hand, CP-odd Higgs fermion interactions could manifest 
already at the tree level, being naturally more sensitive to new physics~\cite{Ellis:2013yxa,Buckley:2015vsa,Boudjema:2015nda,
Mileo:2016mxg,Gritsan:2016hjl,Berge:2008wi,Harnik:2013aja,Dolan:2016qvg,Santos:2015dja,Goncalves:2016qhh,Demartin:2014fia,
Chien:2015xha,Cirigliano:2016nyn,Han:2016bvf,Khatibi:2014bsa,Hagiwara:2016zqz,Kobakhidze:2014gqa,BhupalDev:2007ftb,
Hagiwara:2017ban}. Of special interest is the Higgs coupling to top 
quarks, as  $y_{t}^{SM}\sim 1$. 

Relevant constraints to the CP-structure of the top-Higgs coupling can be indirectly probed via loop-induced interactions in
electric dipole moment (EDM) experiments and gluon fusion $hjj$ production at the LHC~\cite{Brod:2013cka,DelDuca:2006hk,
Englert:2012xt,Dolan:2014upa}. While electron and neutron EDM can set very stringent bounds on CP-mixed top Yukawa, it critically 
assumes the Yukawa coupling with the first generation fermions the same as in the SM, and that new CP-violating interactions are
 limited to the third generation. 
A minor modification on the strength and CP-structure of the Higgs interactions to first generation can considerably degrade these 
constraints~\cite{Brod:2013cka}. Similarly, possible new physics loop-induced contributions can spoil the measurement through 
gluon fusion  $hjj$ production~\cite{Banfi:2013yoa,Azatov:2013xha,Grojean:2013nya,Schlaffer:2014osa,Buschmann:2014twa,Buschmann:2014sia}. Therefore, the direct measurement of this coupling is required to disentangle potential additional new physics effects. 

Analogously to the \emph{direct} (model independent) measurement of the top Yukawa strength, the \emph{direct} measurement 
for its  CP-phase also has the $pp\rightarrow t\bar{t}h$ channel as its  most natural path. Going beyond the signal strength analysis
for this channel becomes even further motivated given  $i)$  the  recent CMS result, showing observation for the $t\bar{t}h$ signal with
5.2$\sigma$ observed (4.2$\sigma$ expected)~\cite{ATLAS-CONF-2017-077,Sirunyan:2018hoz}; and $ii)$  the High-Lumi LHC (HL-LHC) projections,
indicating that this channel will be measured with a very high precision, $\delta y_t <10\%$~\cite{CMS:2013xfa}. Hence, that is the 
approach which we follow in the present study, exploring the spin correlations in the  top pair decays.

The different Higgs-top CP-structure affects the top-spin correlation, that can propagate to the top quark decay products. 
The most natural channel to 
perform such a study is the dileptonic top decay, as the spin analyzing power for charged leptons  is maximal. Spin correlations can be 
enhanced looking at the $t\bar{t}$ rest frame, however the large experimental uncertainties at hadron collider due to top reconstruction
and frame change make this measurement challenging. We will present a method for the top reconstruction that will address these issues,
allowing the construction of relevant CP observables at  $t\bar{t}$ rest frame.

The aim of this paper is twofold. First, we will study direct Higgs-top CP measurement via the $ t \bar t h$ production, exploiting full kinematic 
reconstruction in the dilepton channel. For this purpose we adopt a kinematic reconstruction method presented in Ref. \cite{Debnath:2017ktz}. 
Second, since this reconstruction method was studied only at the parton-level, we would like to investigate its performance further beyond the parton-level, including more 
realistic effects such as parton-shower, hadronization and detector resolution. Although this reconstruction method was initially presented for the 
top quark pair production $ t \bar t$, we will show that it can be easily adopted to the $ t \bar t h$ channel.

This paper is structured as follows. In section~\ref{sec:model}, we will present our setup and the kinematic observables to 
access the CP-phase. In section~\ref{sec:reco}, we will discuss the method for kinematic reconstruction of the dileptonic tops.
In section~\ref{sec:parton}, we show that the angular correlations can be obtained via this method, presenting the results at the 
parton level, while in section \ref{sec:detector}, we perform a full signal and background analysis, including parton-shower, hadronization and detector effects, and discuss the prospects of the CP measurement in the  $t\bar{t}h$ channel with dileptonic top-quarks and $h\rightarrow b\bar{b}$ decays.

\section{Setup and angular observables\label{sec:model}}

We start with the following Lagrangian containing the top Yukawa coupling
\bea
\label{eq:Lag} 
\mathcal{L} &\supseteq& - \frac{m_t}{v} \, K \, \bar{t}  \, ( \cos \alpha + i \gamma_{5} \sin \alpha ) \, t \, h\, , 
\eea
where $v = 246$~GeV is the SM Higgs vacuum expectation value, $K$ is a real number and $\alpha$ represents the Higgs-top 
CP-phase. Hence, the SM Higgs-top interaction is represented by the pure CP-even coupling $(K,\alpha)=(1,0)$, while $(K,\alpha=\pi/2)$ parametrizes a pure CP-odd Higgs boson.

Various observables have been explored in the literature to access the Higgs-top CP-phase in $t\bar{t}h$ events, {\it e.g}., total cross-section, transverse Higgs momentum,  invariant $t\bar{t}$ mass, and spin correlations in the top quark decay products~\cite{Ellis:2013yxa,Buckley:2015vsa,Boudjema:2015nda,Mileo:2016mxg,Gritsan:2016hjl,Harnik:2013aja,Berge:2008wi,Dolan:2016qvg,Santos:2015dja,Goncalves:2016qhh}. The latter is specially
interesting as it  can accurately  probe the  Higgs-top interaction, exploring the spin polarization of the $t\bar{t}$ pair via a shape analysis.

\begin{figure}[t!]
\centering
\includegraphics[scale=0.275]{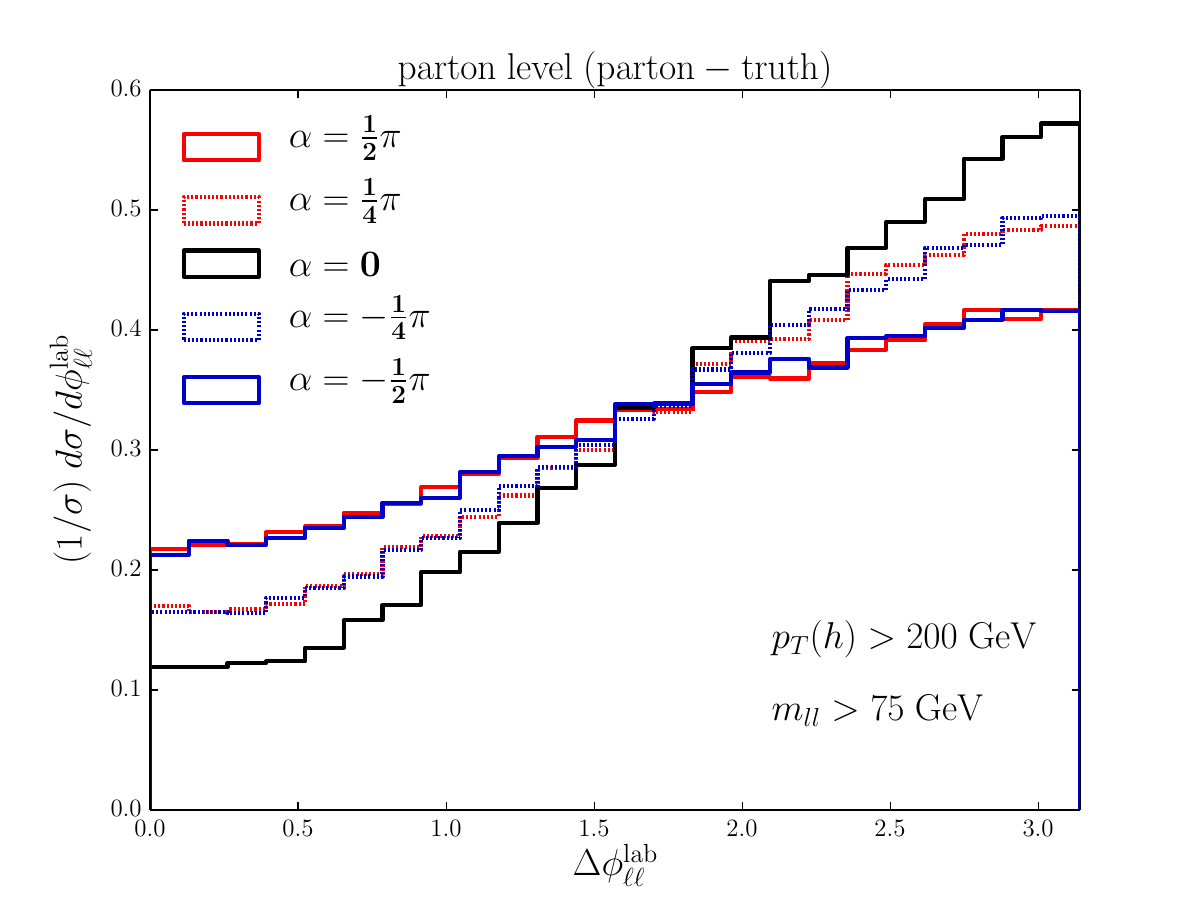}  \hspace{-0.7cm}
\includegraphics[scale=0.275]{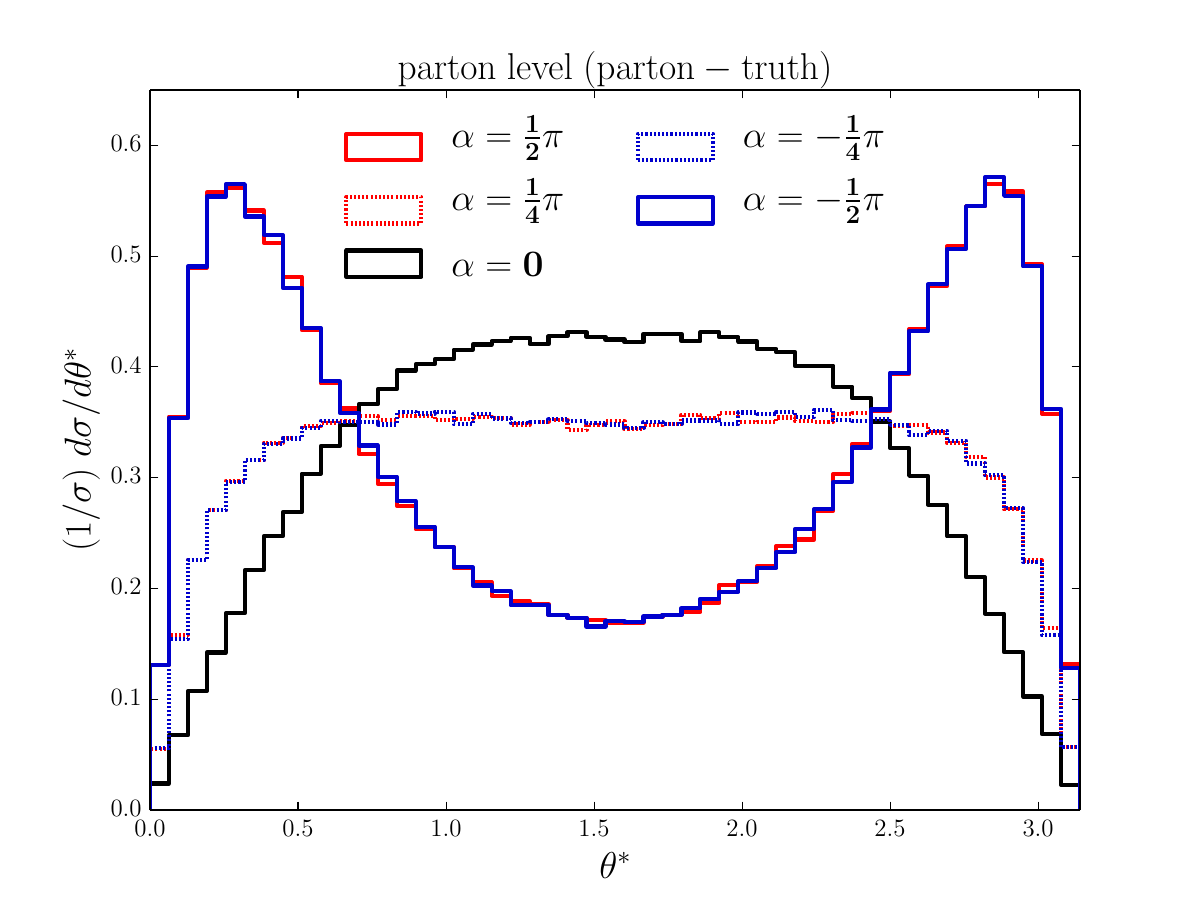}  \hspace{-0.5cm}
\includegraphics[scale=0.285]{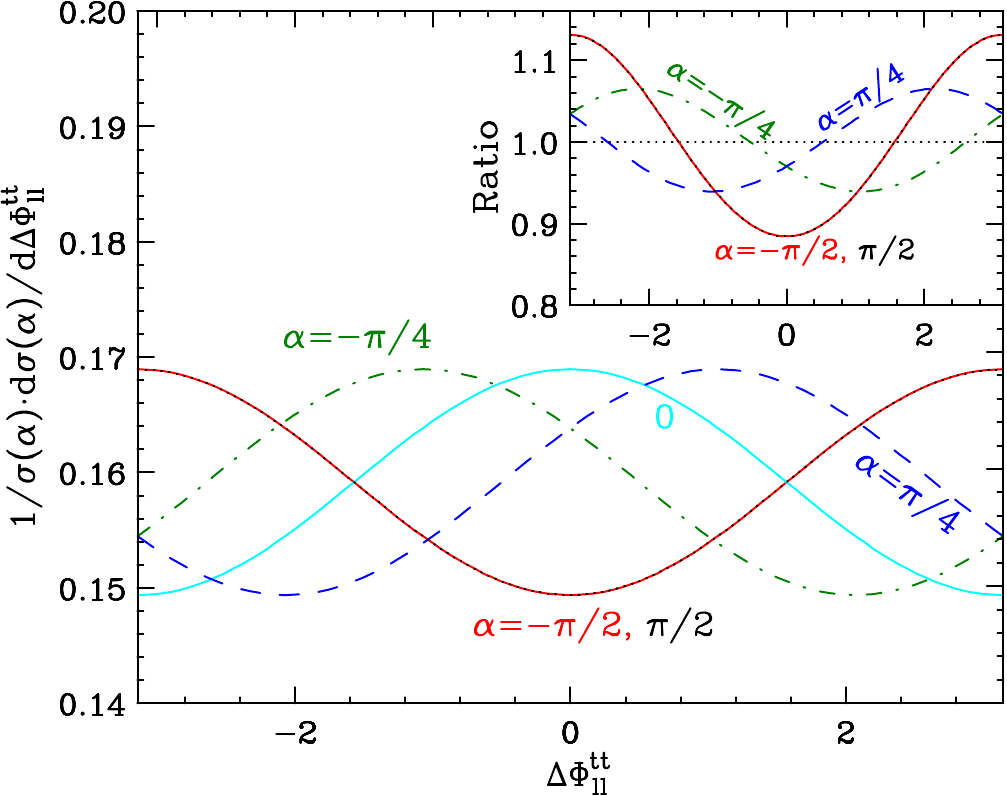} 
\caption{\label{fig:Delta_Phi_parton} \baselineskip 3.0ex 
Left: $\Delta \phi_{\ell \ell}^{\rm lab}$ distribution between the two leptons from the $t \bar t$ decay in the laboratory frame after $p_T (h) > 200 \; \rm{GeV}  $ and $ m_{\ell \ell} > 75 \; \rm{GeV} $ selections.
Middle: Distribution of a collision angle ($\theta^\ast$) of the top with respect to a beam axis in the $t\bar{t}$ rest frame. 
Right: $\Delta \phi_{\ell \ell}^{t\bar t}$ distribution between the two leptons  in the $t\bar{t}$ rest frame for $t\bar{t}h$ production. We display the SM $\alpha=0$ and beyond the SM scenarios with $\alpha=\pm\pi/4,\pm\pi/2$. The ratios of the different hypotheses to the SM are shown in the sub-figure (top right). 
The results are at the parton-truth level, fully reconstructing the particles' momenta at the 14 TeV LHC.
}
\end{figure}

While at hadron colliders the top quarks are unpolarized, the top and anti-top pair are highly correlated. This fact can be experimentally  revealed by spin correlations between the top decay products~\cite{Mahlon:1995zn}. The top-quark spin polarization is transferred to the top decays,  $t\rightarrow W^+ b$ with $W^+\rightarrow \ell^+\nu$ or ${\bar{d}+u}$, where the spin analyzing power is maximal for the charged lepton $\ell^+$ and the down quark $\bar{d}$. Exploring this, Ref.~\cite{Buckley:2015vsa} demonstrates that the difference in azimuthal angle  between the leptons $\Delta \phi_{\ell \ell}^{\rm lab}$  (from top decays) in the laboratory frame can directly reveal the CP-structure of the Higgs-top  interaction with the sensitivity of the measurement substantially enhanced in the boosted Higgs regime,  as shown in the left panel of Fig.~\ref{fig:Delta_Phi_parton}. This study shows that the Higgs-top coupling strength and the CP structure can be directly probed with achievable luminosity at the HL-LHC, using boosted Higgs substructure in the dileptonic channel.

In the present paper, we would like to include observables in the center-of-mass frame of $t \bar t$ system, exploiting novel kinematic reconstruction methods. Among several distributions studied in the $t\bar t$ differential cross-section measurements, we find that the production angle $\theta^\ast$ in the Collins-Soper reference frame brings an interesting correlation, as shown in Fig.~\ref{fig:Delta_Phi_parton} (middle). This $\theta^\ast$ is a collision angle of the top with respect to a beam axis in the $t \bar{t}$ center-of-mass frame and therefore the two top quarks have equal and opposite momenta, with each making the same angle $\theta^*$ with the beam direction~\cite{ATLAS:2016jct}. See Ref.~\cite{Dolan:2016qvg} for a recent application of a similar observable which probes the spin and parity of a new light resonance.

All these variables, including $\Delta \phi_{\ell \ell}^{\rm lab}$ and $\theta^\ast$, are sensitive only to the square terms $\cos^2\alpha$ and $\sin^2\alpha$ (CP-even observables), providing only an indirect measure of CP-violation, missing the interference term between CP-even and odd couplings, $\cos\alpha\sin\alpha$, that can capture a relative coupling sign. 
To define  CP-odd observables, we have to further explore the spin polarization of the $t\bar{t}$ pair. Remarkably, tensor product relations of the top-pair and the final state particles, that follow from totally antisymmetric expressions 
${\epsilon(p_a,p_b,p_c,p_d)\equiv\epsilon_{\mu\nu\rho\sigma}p_a^{\mu}p_b^{\nu}p_c^{\rho}p_d^{\sigma}~}$ (with $\epsilon_{0123}=1$), are examples of such observables. 

In the present work, we will focus on a relevant tensor product that has information on the top and anti-top and the charged leptons from
top-quark decays, maximizing the spin analyzing power: $\epsilon(p_t,p_{\bar{t}},p_{\ell^+},p_{\ell^-})$. In general, this expression leads 
to several terms, making it difficult to define an observable that extracts all its information. However, this relation opportunely simplifies at the
$t\bar{t}$ center of mass (CM) frame, resulting in a single triple product 
\bea
\label{eq:obs1} 
\epsilon(p_t,p_{\bar{t}},p_{\ell^+},p_{\ell^-})|_{t\bar{t}~CM}  \propto  p_t \cdot  (p_{\ell^+} \times p_{\ell^-} ) \;,
\eea
provided that we can fully reconstruct the $t\bar{t}$ CM frame. We further explore this relation to define our CP-odd observable 
\bea
\label{eq:obs2} 
\Delta \phi_{\ell \ell }^{t \bar t} =  \text{sgn}\left[\vec{p}_t \cdot  (\vec{p}_{\ell^+} \times \vec{p}_{\ell^-} )\right] 
\arccos\left[\frac{\vec{p}_t\times \vec{p}_{\ell^+}}{|\vec{p}_t\times \vec{p}_{\ell^+}|}\cdot \frac{\vec{p}_t\times \vec{p}_{\ell^-}}{|\vec{p}_t\times \vec{p}_{\ell^-}|}\right] \;,
\eea
that is defined in the $[-\pi,\pi]$ range. In Fig.~\ref{fig:Delta_Phi_parton} (right), we display the $\Delta \phi_{\ell \ell }^{t\bar t}$ distributions at 
the parton-truth level for different CP hypotheses $\alpha$. The CP-mixed cases $\alpha=\pi/4$ from $-\pi/4$ display 
different distribution shapes, confirming that $\Delta \phi_{\ell \ell }^{t\bar t }$ is a truly CP-odd observable.  

One may quantify these differences via an asymmetry,  comparing the number of events with  positive and negative $\Delta \phi^{t \bar{t}}_{\ell \ell }$ \cite{Mileo:2016mxg}: 
\bea
\label{eq:A1} 
\mathcal{A}_{\ell\ell} &\equiv& \frac{  N( \Delta \phi^{t \bar{t}}_{\ell\ell } >0) - N( \Delta \phi^{t \bar{t}}_{\ell\ell } < 0 )   }{  N( \Delta \phi^{t \bar{t}}_{\ell\ell } >0) + N( \Delta \phi^{t \bar{t}}_{\ell\ell } < 0 ) } \, ,
\eea
where $\mathcal{A}_{\ell\ell}\in [-1,1]$. 
While the asymmetry $\mathcal{A}_{\ell\ell}$ results in deviations from the SM hypothesis of at maximum $\mathcal{O}(4\%)$ 
(for $\alpha \approx \pm \frac{\pi}{4}$), $\Delta \phi_{\ell \ell}^{t\bar t}$ presents parameter space regions that can reach up to $\mathcal{O}(10\%)$ of difference in ratio $ \left (  \frac{1}{\sigma(\alpha)}\cdot\frac{d\sigma(\alpha)}{d\Delta\phi_{\ell\ell}^{t\bar t}} \right ) / \left (  \frac{1}{\sigma(0)} \cdot\frac{d\sigma(0)}{d\Delta\phi_{\ell\ell}^{t\bar t}} \right )$, as shown in the subfigure of the right plot. The latter leads to a potentially stronger distinguishing power that can be explored via a shape analysis. Due to difficulty in event reconstruction to go to the $t\bar t$ rest frame, the $\Delta\phi_{\ell \ell}^{t\bar t}$ observable has not been investigated in a realistic analysis so far. 
In this study, we shall attempt to reconstruct the $\theta^\ast$ and $\Delta \phi_{\ell \ell }^{t \bar t}$ variable at hadron-level including detector resolution. 
We will then examine how these two observables ($\Delta\phi_{\ell \ell}^{t\bar t}$ and $\theta^\ast$) would improve the existing analysis with the laboratory angle ($\Delta \phi_{\ell \ell}^{\rm lab}$). We will make a brief comment on the sign of CP angle as well. 

\section{Brief review of kinematic reconstruction\label{sec:reco}}

\begin{figure}[t!]
\centering
\includegraphics[scale=0.85]{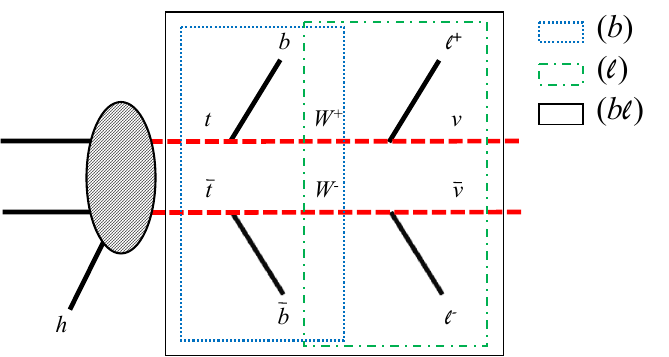}
\caption{The event topology considered in this paper, 
together with the three possible subsystems. 
The blue dotted, the green dot-dashed, and
  the black solid boxes indicate the subsystems $(b)$, $(\ell)$, and
  $(b\ell)$, respectively. \label{fig:decaysubsystem}}
\end{figure}

In this section, we briefly review the reconstruction method that we adopt. Our algorithm is entirely based on mass minimization. Thus, it is more flexible  for  new physics analyses  and  robust for our  spin-correlation study~\footnote{See Refs. \cite{Lester:1999tx,Konar:2009wn,Burns:2008va,Konar:2009qr} for $M_{T2}$ and its various extensions and Refs. \cite{Barr:2011xt,Cho:2014naa,Cho:2015laa,Konar:2008ei,Konar:2010ma} for four dimensional variables. We refer to Refs. \cite{Barr:2010zj,Barr:2011xt} for reviews on various kinematic variables.}. 
The event topology considered in this paper is shown in Fig. \ref{fig:decaysubsystem}, together with three possible subsystems. 
The blue dotted, the green dot-dashed, and the black solid boxes indicate the subsystems $(b)$, $(\ell)$, and $(b\ell)$, respectively. We consider that the Higgs (denoted as $h$) is fully reconstructed, in which case the only source of the missing transverse momentum is two neutrinos from the top decays.

In the presence of two missing particles at the end of a cascade decay, 
$M_{T2}$ provides a good estimate of mass information in the involved decay \cite{Lester:1999tx,Barr:2010zj,Burns:2008va,Barr:2011xt}.
Following notations and conventions of Ref. \cite{Debnath:2017ktz}, we define $M_{T2}$ as follows: 
\bea
M_{T2} (\tilde m) &\equiv& \min_{\vec{q}_{1T},\vec{q}_{2T}}\left\{\max\left[M_{TP_1}(\vec{q}_{1T},\tilde m),\;M_{TP_2} (\vec{q}_{2T},\tilde m)\right] \right\} \, ,
\label{eq:mt2def}\\
\vec{q}_{1T}+\vec{q}_{2T} &=& \mpt  \;,  \nonumber
\eea 
where $M_{TP_i}$ ($i=1, 2$) is the transverse mass of the decaying particle in the $i$-th side and $\tilde m$ is a test mass, which we set to $zero$ in our study. $\vec{q}_{iT}$ is the unknown transverse momentum of the $i$-th missing particle, which is a neutrino in this case. Individual values ($\vec{q}_{1T}$ and $\vec{q}_{2T}$) are unknown and only their sum ($\vec{q}_{1T}+\vec{q}_{2T}$) is constrained by the total missing transverse momentum, $\mpt $.

Another mass-constraining variable is the $M_N$ \cite{Barr:2011xt,Debnath:2017ktz,Kim:2017awi}, which is the (3+1)-dimensional version of Eq.~(\ref{eq:mt2def}): 
\bea
M_{2} (\tilde m) &\equiv& \min_{\vec{q}_{1},\vec{q}_{2}}\left\{\max\left[M_{P_1}(\vec{q}_{1},\tilde m),\;M_{P_2} (\vec{q}_{2},\tilde m)\right] \right\} \, ,
\label{eq:m2def}\\
\vec{q}_{1T}+\vec{q}_{2T} &=& \mpt  \;, \nonumber
\eea 
where the {\em actual} parent masses ($M_{P_i}$) are considered instead of their transverse masses ($M_{T P_i}$).
Note that the minimization is now performed over the 3-component momentum vectors $\vec{q}_{1}$ and $\vec{q}_{2}$ \cite{Barr:2011xt}.
In fact, at this point the two definitions (\ref{eq:mt2def}) and (\ref{eq:m2def}) are equivalent, in the sense that the resulting two variables, $M_{T2}$ and $M_2$, will have the same numerical value \cite{Ross:2007rm,Barr:2011xt,Cho:2014naa}. 

However, $M_2$ begins to differ from $M_{T2}$ when applying additional kinematic constraints 
beyond the missing transverse momentum condition $\vec{q}_{1T}+\vec{q}_{2T} = \mpt $. Then, the $M_2$ variable can be further refined and one can obtain non-trivial variants as shown below \cite{Cho:2014naa}:
\bea
M_{2XX} &\equiv& \min_{\vec{q}_{1},\vec{q}_{2}}\left\{\max\left[M_{P_1}(\vec{q}_{1},\tilde m),\;M_{P_2} (\vec{q}_{2},\tilde m)\right] \right\},  
\label{eq:m2XXdef}
\\
\vec{q}_{1T}+\vec{q}_{2T} &=& \mpt   \nonumber
\eea 
\bea
M_{2CX} &\equiv& \min_{\vec{q}_{1},\vec{q}_{2}}\left\{\max\left[M_{P_1}(\vec{q}_{1},\tilde m),\;M_{P_2} (\vec{q}_{2},\tilde m)\right] \right\},  
\label{eq:m2CXdef}\\
\vec{q}_{1T}+\vec{q}_{2T} &=& \mpt   \nonumber \\
M_{P_1}&=& M_{P_2} \nonumber 
\eea 
\bea
M_{2XC} &\equiv& \min_{\vec{q}_{1},\vec{q}_{2}}\left\{\max\left[M_{P_1}(\vec{q}_{1},\tilde m),\;M_{P_2} (\vec{q}_{2},\tilde m)\right] \right\},  
\label{eq:m2XCdef}\\
\vec{q}_{1T}+\vec{q}_{2T} &=& \mpt   \nonumber \\
M_{R_1}^2&=& M_{R_2}^2 \nonumber 
\eea 
\bea
M_{2CC} &\equiv& \min_{\vec{q}_{1},\vec{q}_{2}}\left\{\max\left[M_{P_1}(\vec{q}_{1},\tilde m),\;M_{P_2} (\vec{q}_{2},\tilde m)\right] \right\}.
\label{eq:m2CCdef}\\
\vec{q}_{1T}+\vec{q}_{2T} &=& \mpt   \nonumber \\
M_{P_1}&=& M_{P_2} \nonumber  \\
M_{R_1}^2&=& M_{R_2}^2 \nonumber 
\eea 
Here $M_{P_i}$ ($M_{R_i}$) is the mass of the parent (relative) particle in the $i$-th decay chain and a subscript ``$C$" 
indicates that an equal mass constraint is applied for the two parents (when ``$C$" is in the first position) or for the relatives
(when ``$C$" is in the second position). A subscript ``$X$" simply means that no such constraint is applied.
Note that $M_{2XX}$ in Eq. (\ref{eq:m2XXdef}) is the same as the original definition of $M_2$ in Eq. (\ref{eq:m2def}) and the subscript $(XX)$ is added explicitly to indicate that no extra constraints are imposed during the minimization. 
In any given subsystem ($(b)$, $(\ell)$ or $(b\ell)$), these variables (\ref{eq:mt2def}-\ref{eq:m2CCdef}) are related as follows \cite{Cho:2014naa}
\bea
M_{T2} = M_{2XX}=M_{2CX} \leq M_{2XC} \leq M_{2CC}. 
\label{eq:hierarchy}
\eea

More specifically, in the $t \bar t$-like production ($t\bar t + X$ where $X$ is fully reconstructed), we could use the experimentally measured $W$-boson mass, $m_W$, and introduce the following variable: 
\bea
M_{2CW}^{(b\ell)} &\equiv& \min_{\vec{q}_{1},\vec{q}_{2}}\left\{\max\left[M_{t_1}(\vec{q}_{1},\tilde m),\;M_{t_2} (\vec{q}_{2},\tilde m)\right] \right\} \, .
\label{eq:m2CWdef}\\
\vec{q}_{1T}+\vec{q}_{2T} &=& \mpt   \nonumber \\
M_{t_1}&=& M_{t_2} \nonumber  \\
M_{W_1}&=& M_{W_2} = m_W \nonumber 
\eea 
Similarly, taking the mass $m_t$ of the top quark in the minimization, we can define a new variable in the $(\ell)$ subsystem:
\bea
M_{2Ct}^{(\ell)} &\equiv& \min_{\vec{q}_{1},\vec{q}_{2}}\left\{\max\left[M_{W_1}(\vec{q}_{1},\tilde m),\;M_{W_2} (\vec{q}_{2},\tilde m)\right] \right\}.
\label{eq:m2Ctdef}\\
\vec{q}_{1T}+\vec{q}_{2T} &=& \mpt   \nonumber \\
M_{W_1}&=& M_{W_2} \nonumber  \\
M_{t_1}&=& M_{t_2} = m_t \nonumber 
\eea 

Although these mass-constraining variables are proposed for mass measurement originally, 
one could use them for other purposes such as measurement of spins and couplings \cite{Baringer:2011nh}. In our study, we use these variables to fully reconstruct the final state of our interest, with the unknown momenta obtained via minimization procedure. These momenta may or may not be true particle momenta but they provide important non-trivial correlations with other visible particles in the final state, which helps reconstruction. 

Based on Ref. \cite{Debnath:2017ktz}, we define the following parameter space:
\beq
\left(x,y,z\right) \equiv
\left(   \,
m_{b\ell}^{max}-\max_i\{m^{(i)}_{b\ell}\}, \,
m_t - M_{2CW}^{(b\ell)}, \,
m_W - M_{2Ct}^{(\ell)} \,
\right) \, ,
\label{setofthreeWt}
\eeq 
where $m^{(i)}_{b\ell}$ is the invariant mass of $b$ and $\ell$ in $i$-th pairing ($i=1,2$), and $m_{b\ell}^{max} = \sqrt{ m_t^2 - m_W^2}$ (in the $m_b \to 0$ limit). Since there are two possible ways of paring $b$ and $\ell$ in the dilepton channel of the $t \bar t$-like events, we repeat the same calculation for each partitioning. 
Then the correct combination would respect the anticipated end points of $m_{b\ell}$, $M_{2CW}^{(b\ell)}$ and $M_{2Ct}^{(\ell)}$, leading to positive $x$, $y$, and $z$.
On the other hand, the wrong pairing could give either sign. Finally, by requiring that the partition which gives more ``plus" sign as the ``correct" one, we can resolve two-fold ambiguity. Then we treat the corresponding momenta of two missing particles (which are obtained via the minimization procedure) as ``real'' momenta of two missing neutrinos. 
If both partitions give the same numbers of positive and negative signs (called ``unresolved case''), we discard those events. 
From Ref. \cite{Debnath:2017ktz}, the efficiency of this method is known to be about 88\%, including unresolved events with a coin flip, 50\% probability of picking the right combination. Since we ignore those events to obtain a high-purity sample, the corresponding efficiency becomes 83\%. 
We also note that we assign the negative sign for a partitioning, if a viable solution is not found during minimization. This is because the wrong pairing would fail more often than the correct paring.  
With the obtained neutrino momenta, now we can reconstruct momenta of $W$s and top quarks for the CP measurement of the top-Yukawa coupling.

\section{Top-Higgs Yukawa coupling with $M_2$-assisted reconstruction~\label{sec:analysis}}

We show our parton-level results in section \ref{sec:parton}, and 
detector-level (including parton-shower, hadronization,  and detector resolution for signal and backgrounds) in section \ref{sec:detector}. 
For our parton-level study, we assume that the Higgs is fully reconstructed. 
We separate these semi-realistic effects to better examine the capability and feasibility of reconstruction methods in the dileptonic $t \bar t h$ production. 
Throughout our study, we use \verb|OPTIMASS| \cite{Cho:2015laa} to obtain momenta of two invisible neutrinos, following the reconstruction method described in the previous section.

\subsection{Parton-level analysis\label{sec:parton}}

Parton level events are generated at leading order by \texttt{MadGraph5\_aMC@NLO}\ \cite{Alwall:2014hca} in chain with \frules~package~\cite{Alloul:2013bka} without any generation level cuts. We use the default \texttt{NNPDF2.3QED} parton distribution function~\cite{Ball:2013hta} with dynamical renormalization and factorization scales set to $m_T^2$ (transverse mass of the visible system) at the 14 TeV LHC. In this section, we focus on comparison between Monte-Carlo truth and parton-level results without worrying about effects of hadronization and parton-shower, which will be the topic in the next section. 
Performing the procedure described in the previous section, we obtain the momenta of two neutrinos and also resolve two fold ambiguity in the dilepton final state, which allows full reconstruction of the final state. No cuts are employed for parton-level studies, unless we mention explicitly.
\begin{figure}[t]
\centering
\includegraphics[scale=0.27]{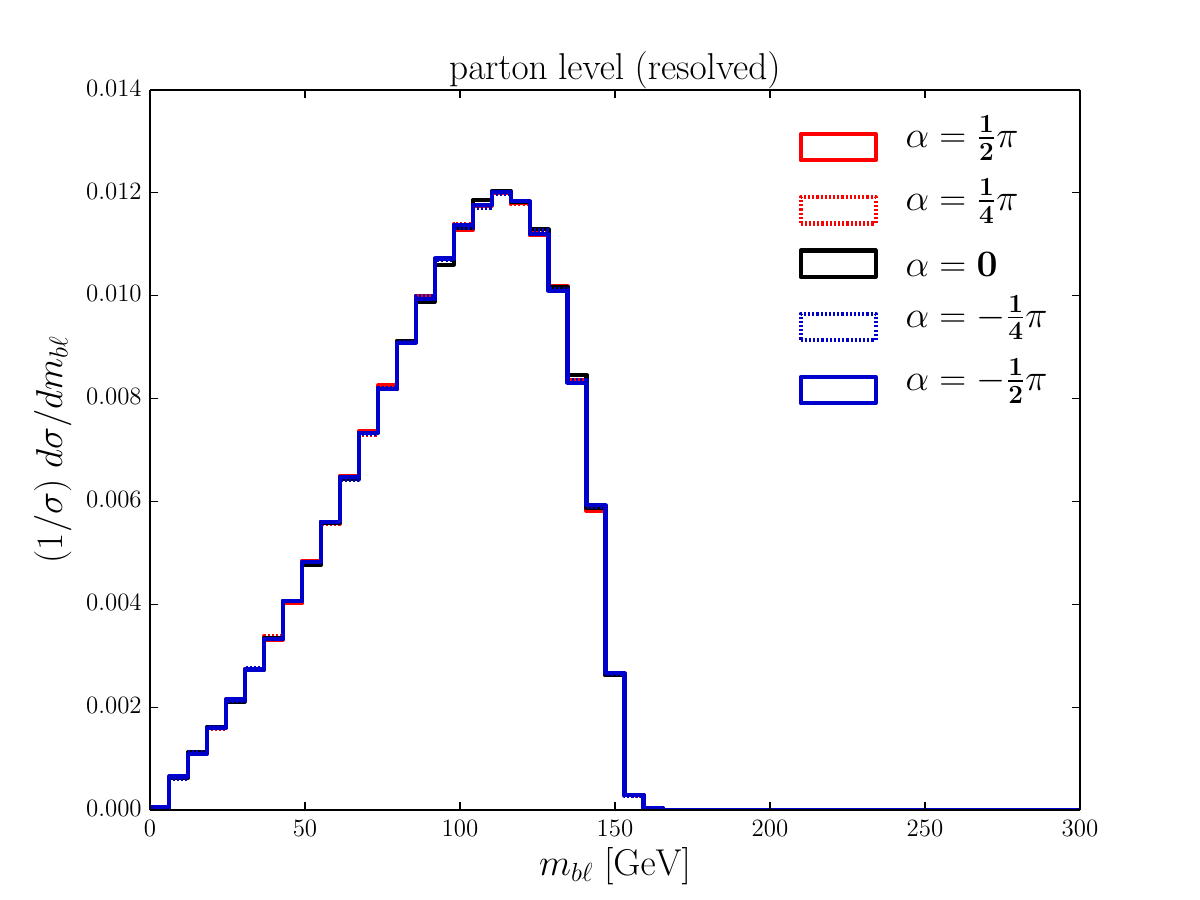} \hspace{-0.7cm}
\includegraphics[scale=0.27]{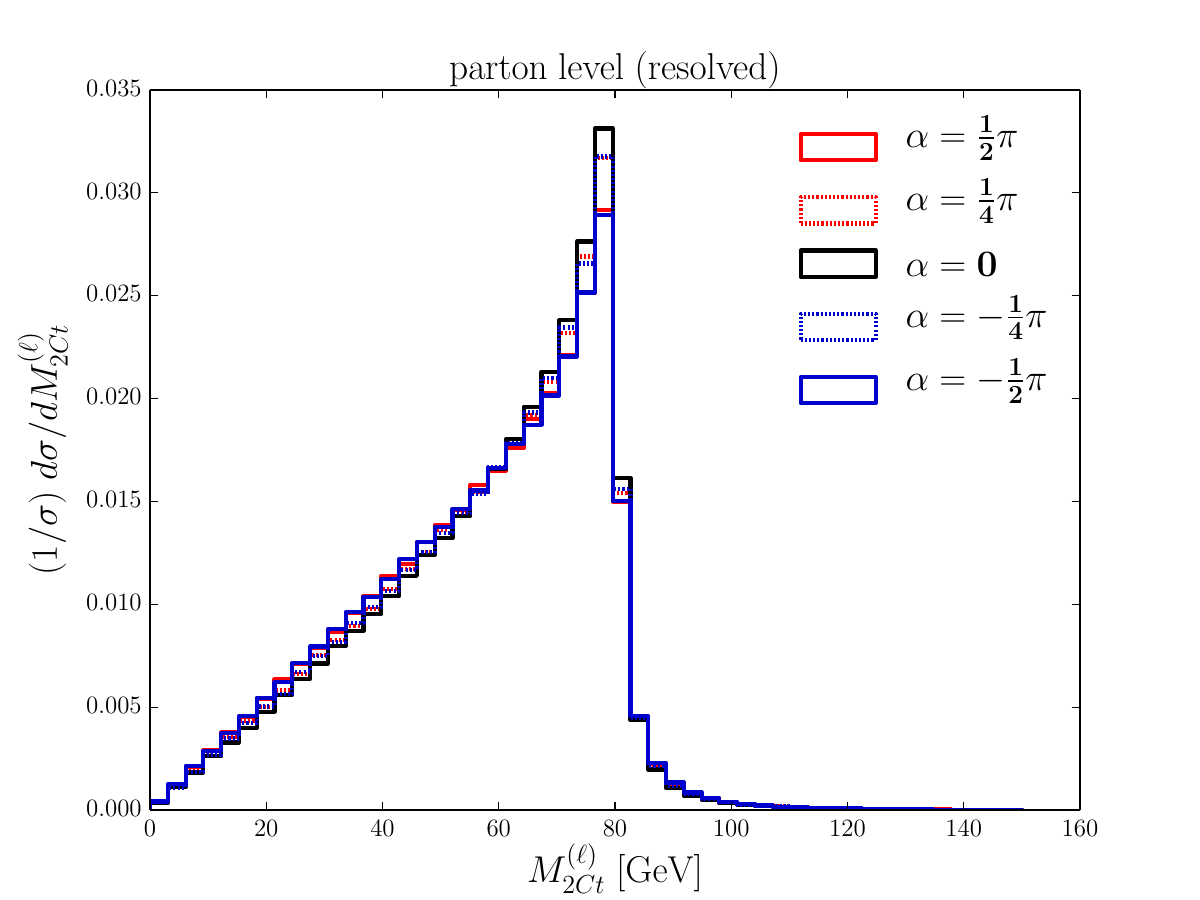}  \hspace{-0.7cm}
\includegraphics[scale=0.27]{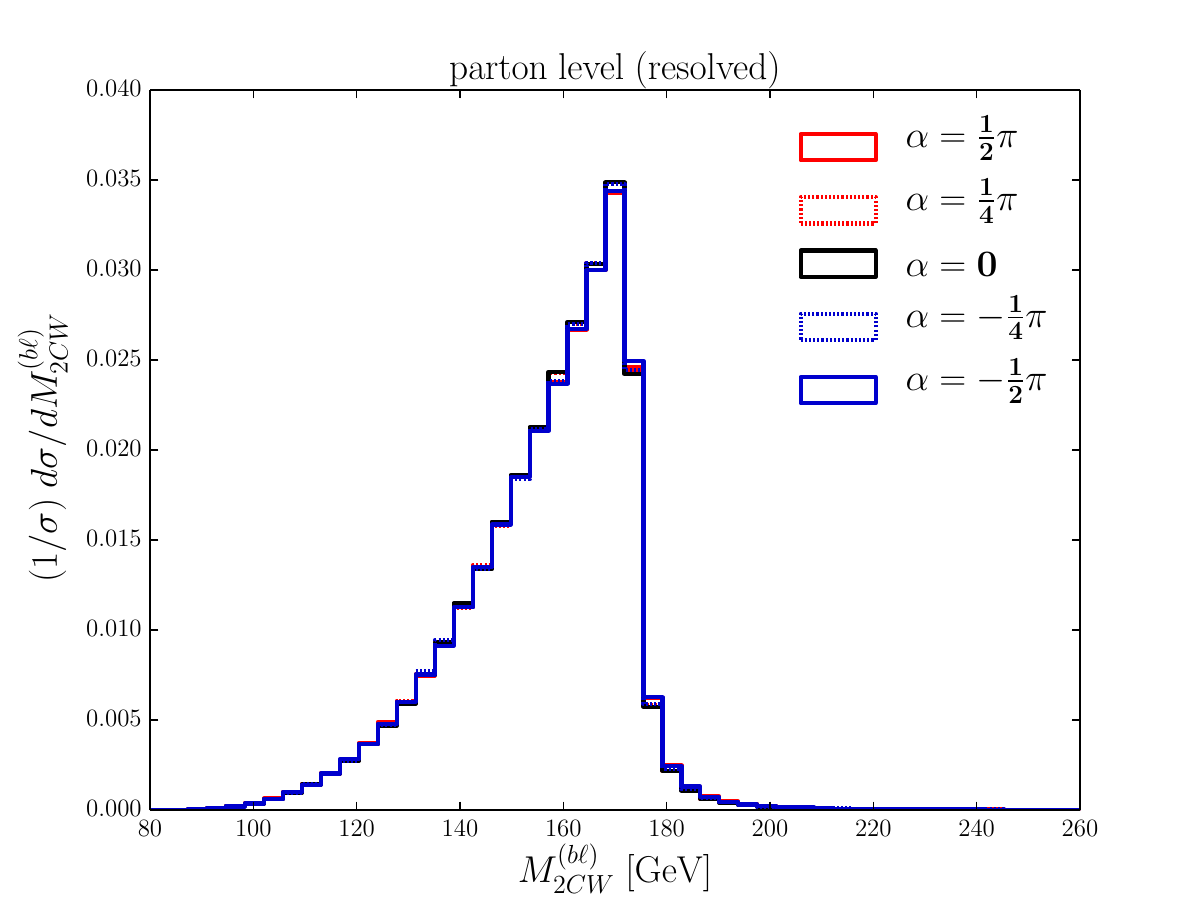} 
\caption{\label{fig:dists} Distributions of $m_{b\ell}$ (left), $M_{2Ct}^{(\ell)}$ (middle), and $M_{2CW}^{(b\ell)}$ (right) for different CP phases. 
}
\end{figure}
Distributions of $m_{b\ell}$, $M_{2Ct}^{(\ell)}$, and $M_{2CW}^{(b\ell)}$ are shown in Fig. \ref{fig:dists} for different values of $\alpha$. Note that, by construction, $M_{2Ct}^{(\ell)}$ and $M_{2CW}^{(b\ell)}$ are bounded above by the mass of the $W$ boson and top quark, $M_{2Ct}^{(\ell)} \leq m_W$ and $M_{2CW}^{(b\ell)} \leq m_t$. 
A small fraction of events which leak beyond the expected mass bounds is due to finite width effects of the top quark and $W$ boson. Also there is small contamination coming from wrong pair of $b$-quark and $\ell$, although the purity of the samples is known to be 96\% \cite{Debnath:2017ktz}.  
Note that, throughout this paper, all plots are generated with the ``resolved'' events, after discarding ``unresolved'' ones. We find that the efficiency of our method is $\epsilon = 81.38 \%$, which is consistent with 83\% as in Ref. \cite{Debnath:2017ktz}.
The resolved events contain both correct and wrong combinations and the fraction of the correct combination out of the resolved events is defined as purity. 

\begin{figure}[t]
\centering
\includegraphics[scale=0.38]{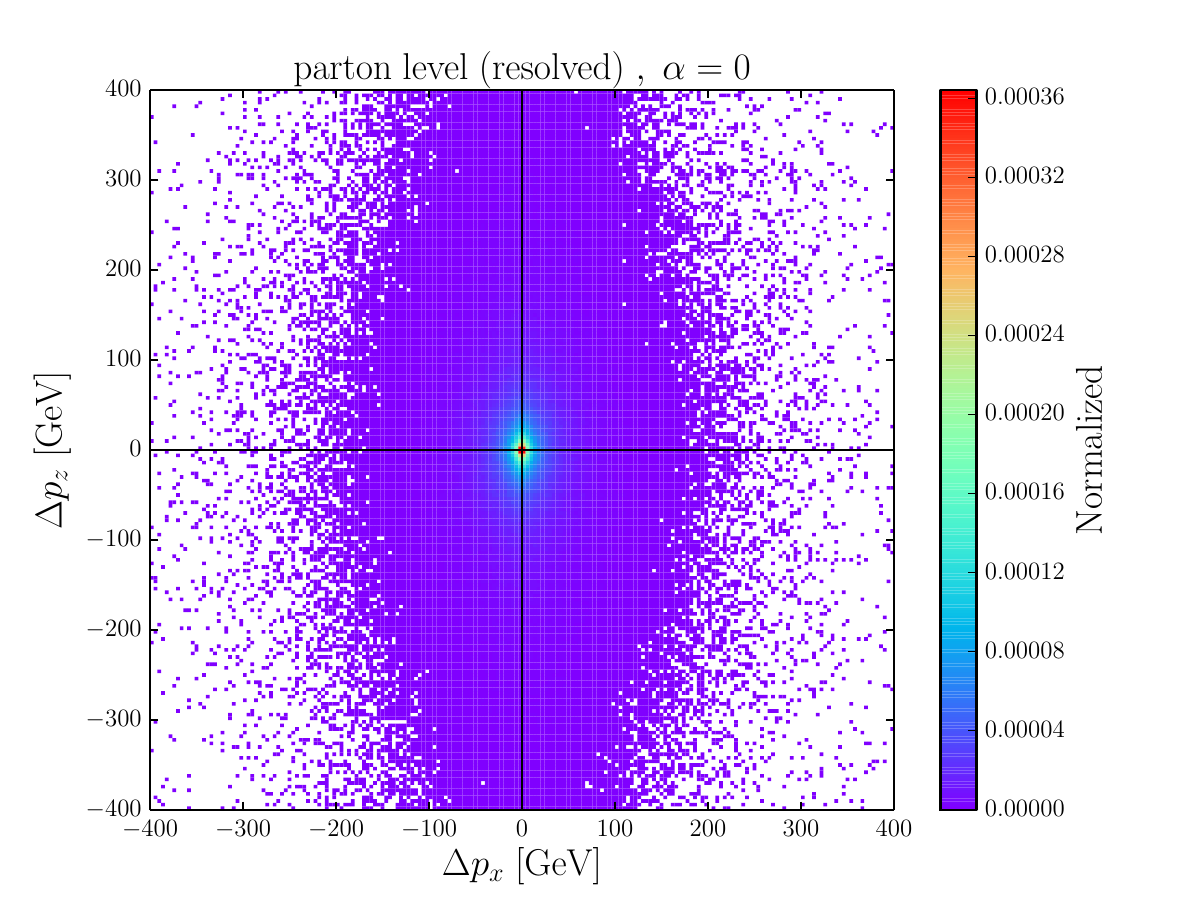} \hspace{-0.7cm} 
\includegraphics[scale=0.38]{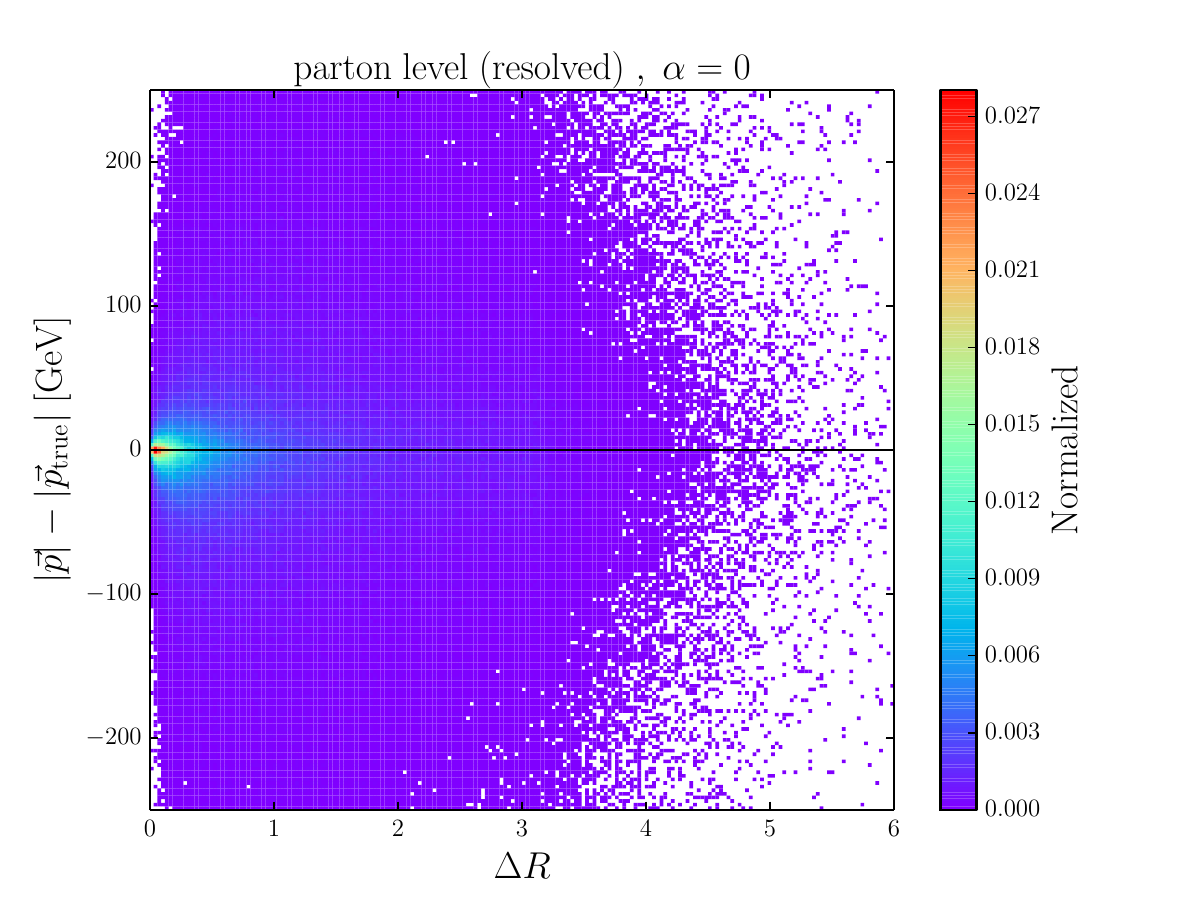}  \\
\includegraphics[scale=0.38]{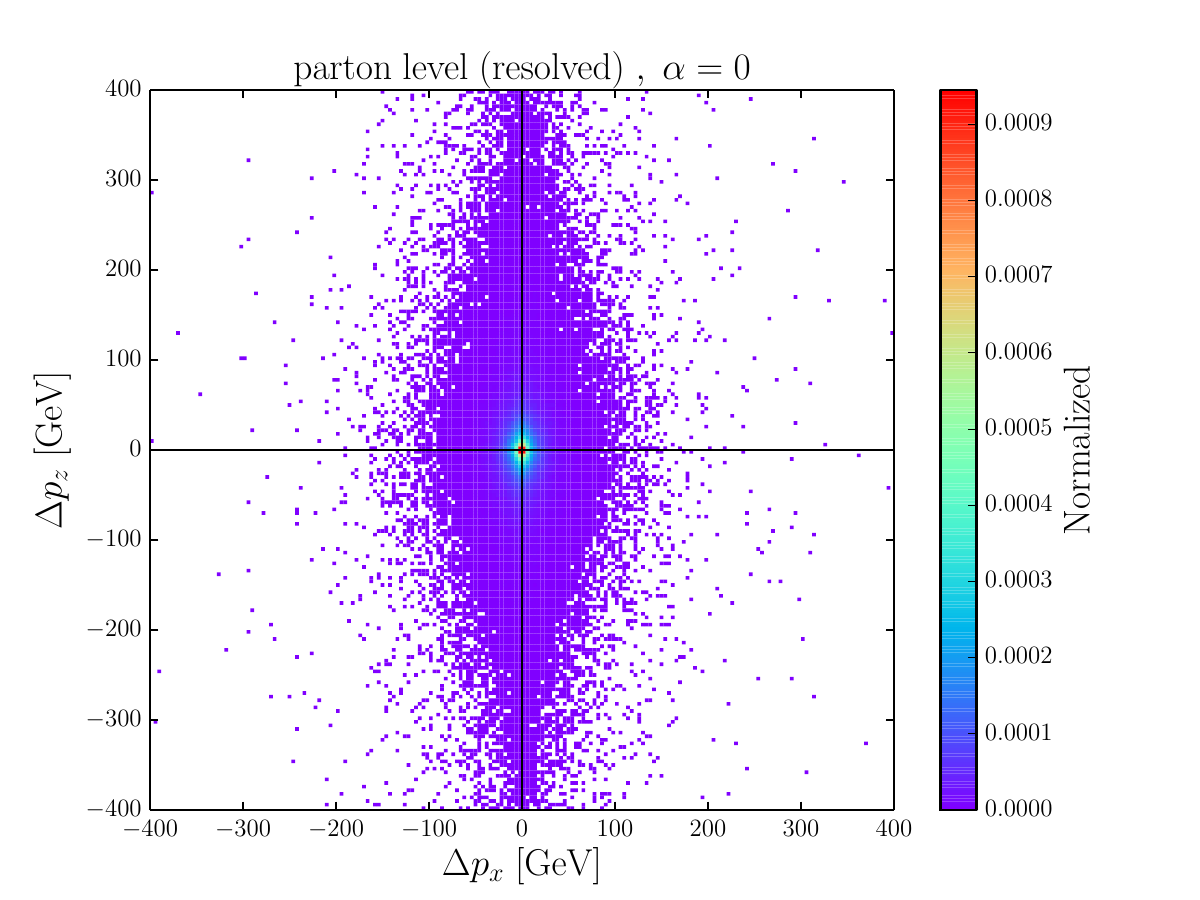} \hspace{-0.7cm} 
\includegraphics[scale=0.38]{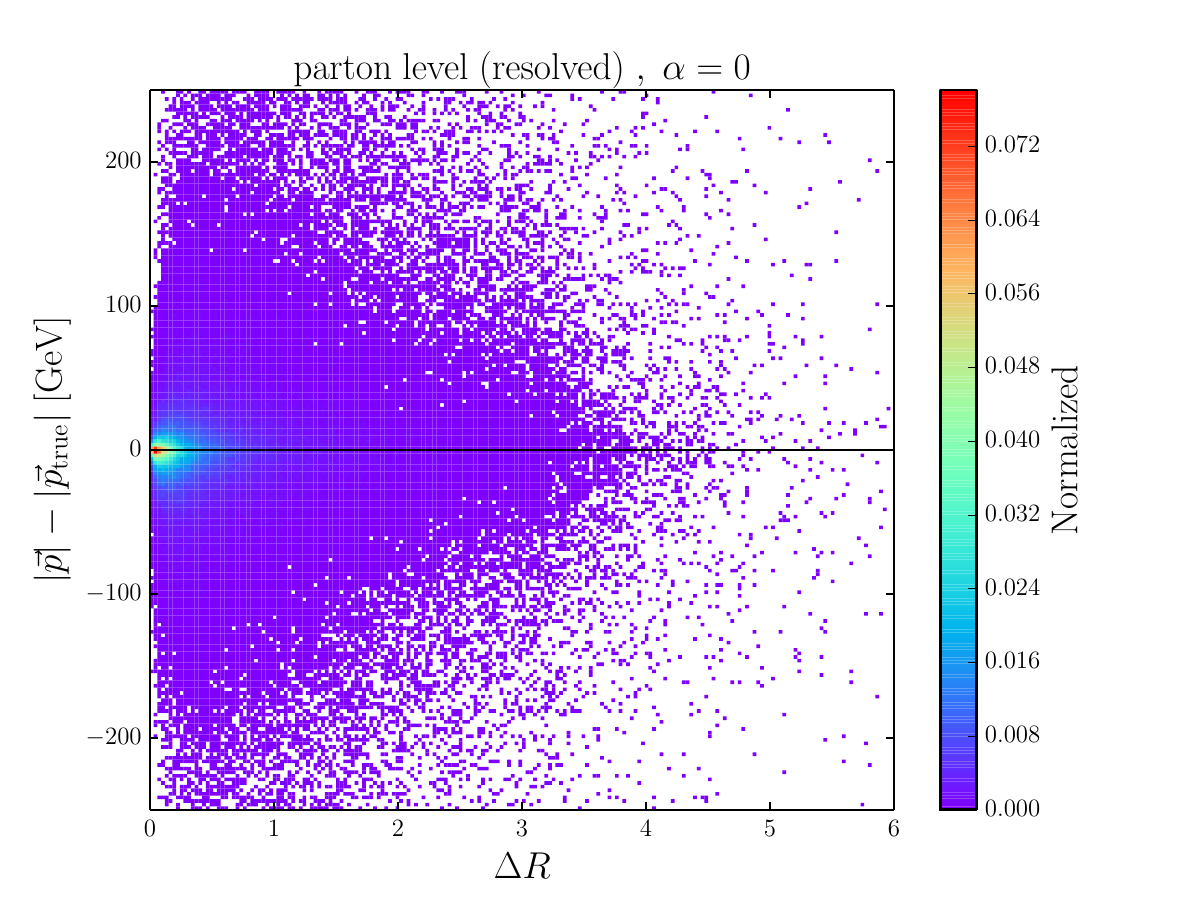} 
\caption{\label{fig:parton_momentum} Correlation between $\Delta p_z$ and $\Delta p_x$, and $| \vec p \; | - | \vec p_{true} |$ and $\Delta R( \vec p ,  \vec p_{true})$ for $M_{2CW}^{(b\ell)}$ (top) without and (bottom) with a mass cut $165 \; {\rm GeV} < M_{2CW}^{(b\ell)} < 175 \; \rm{GeV}$. The corresponding efficiencies are 81.38\% with 96.4\% of purity and 24.86\% with 97.9\% purity, respectively.}
\end{figure}
To examine performance of momentum reconstruction, we show in Fig. \ref{fig:parton_momentum} correlations between $\Delta p_x \equiv p_{x, true} - p_x$ and $\Delta p_z \equiv p_{z, true} - p_z$, and between the difference in magnitude $| \vec p \; | - | \vec p_{true} |$ and the direction mismatch $\Delta R( \vec p ,  \vec p_{true})$ for $M_{2CW}^{(b\ell)}$ for the SM case ($\alpha=0$). Other CP angles show similar results.  
Here $\vec p_{true}$ is the true momentum of a neutrino and $\vec p$ is the momentum from the minimization using \verb|OPTIMASS|. In the upper panel, the scatter plots are generated without any cuts, while a mass cut (165 GeV $< M_{2CW}^{(b\ell)} < 175$ GeV) is applied in the bottom panel, leading to the $\epsilon = 24.86 \%$ efficiency with 97.9\% of purity. 
A relaxed cut, 160~GeV${ < M_{2CW}^{(b\ell)} < 175}$~GeV, gives a slightly higher efficiency $\epsilon$ = 35.82 \% with with 97.7\% of purity. At this point, purity of the resolved sample is already high but the momentum resolution gets improved with a tighter mass cut.
Similar results are expected when using $M_{2Ct}^{(\ell)}$.      

As shown in Ref. \cite{Buckley:2015vsa}, the difference in azimuthal angles of two isolated leptons in the laboratory frame $\Delta \phi^{\rm lab}_{\ell \ell}$ provides a good discrimination of different CP angles at the boosted regime. We reproduce this result as already shown in the left panel of Fig.~\ref{fig:Delta_Phi_parton}.  
Once the cuts of $p_T (h) > 200 \; \rm{GeV}  $ and $ m_{\ell \ell} > 75 \; \rm{GeV} $ are applied, the distributions acquire high distinguishing power, as shown in the figure. Thanks to the fact that it depends only on the leptons, and it is reconstructed at the laboratory frame, this observable displays small uncertainties. 

\begin{figure}[t]
\centering
\includegraphics[scale=0.38]{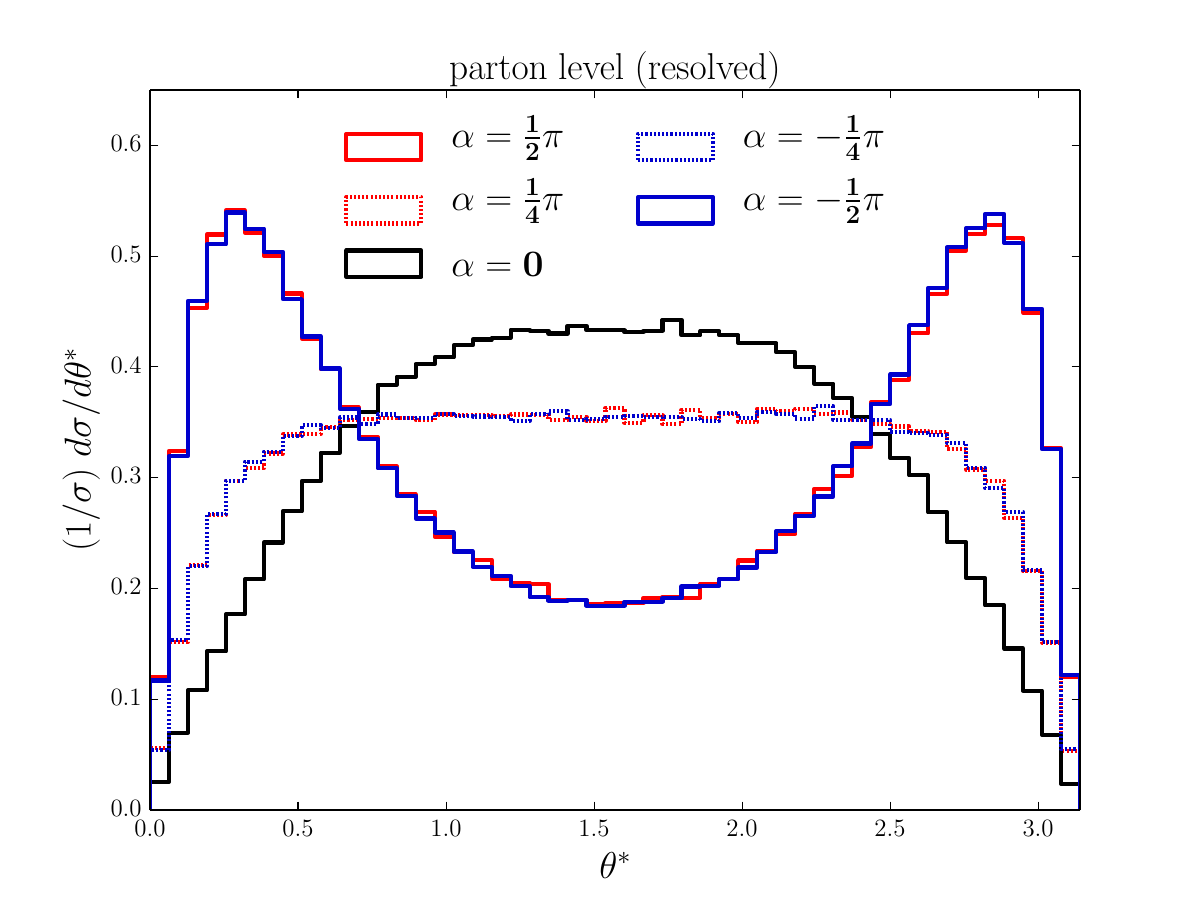} \hspace{-0.72cm} 
\includegraphics[scale=0.38]{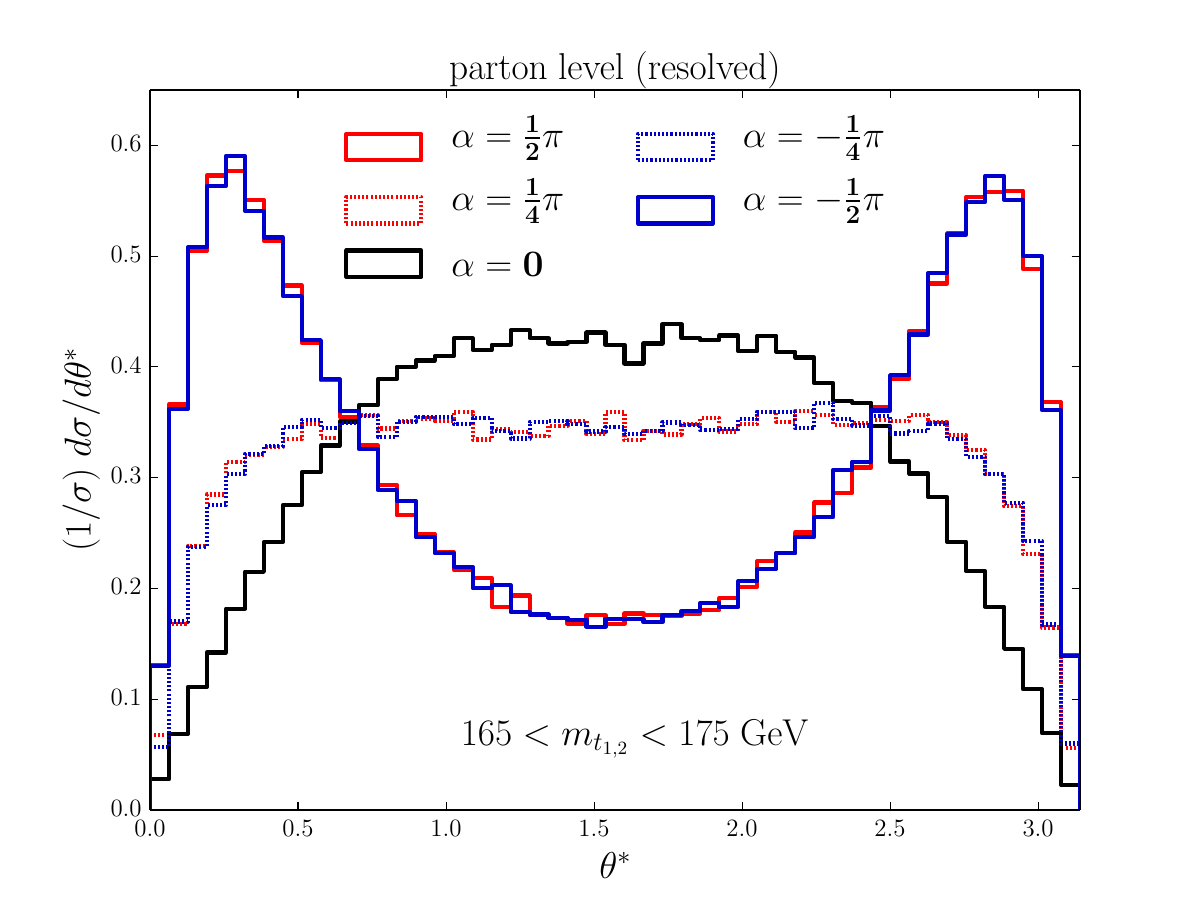} 
\caption{\label{fig:cosparton} \baselineskip 3.0ex 
Distributions of $\theta^\ast$  for various values of $\alpha$ before (left) and after (right) the ${165 \; {\rm GeV} < M_{2CW}^{(b\ell)} < 175 \; \rm{GeV} }$ cut.  
} 
\end{figure}

Having reconstructed full four-momenta of each top, we form $\theta^\ast$ shown in Fig.~\ref{fig:cosparton}, which is the production angle in the Collins-Soper reference frame~\cite{ATLAS:2016jct}. This  distribution exhibits very little sensitivity to the adopted reconstruction procedure  and retains the corresponding shape at Mone-Carlo truth (see the middle plot in Fig.~\ref{fig:Delta_Phi_parton} for comparison). This is partially due to a much simpler structure of $\theta^\ast$ as compared to the shape of other distributions such as $\Delta \phi^{t \bar{t}}_{\ell\ell }$. 

\begin{figure}[t]
\centering
\includegraphics[scale=0.38]{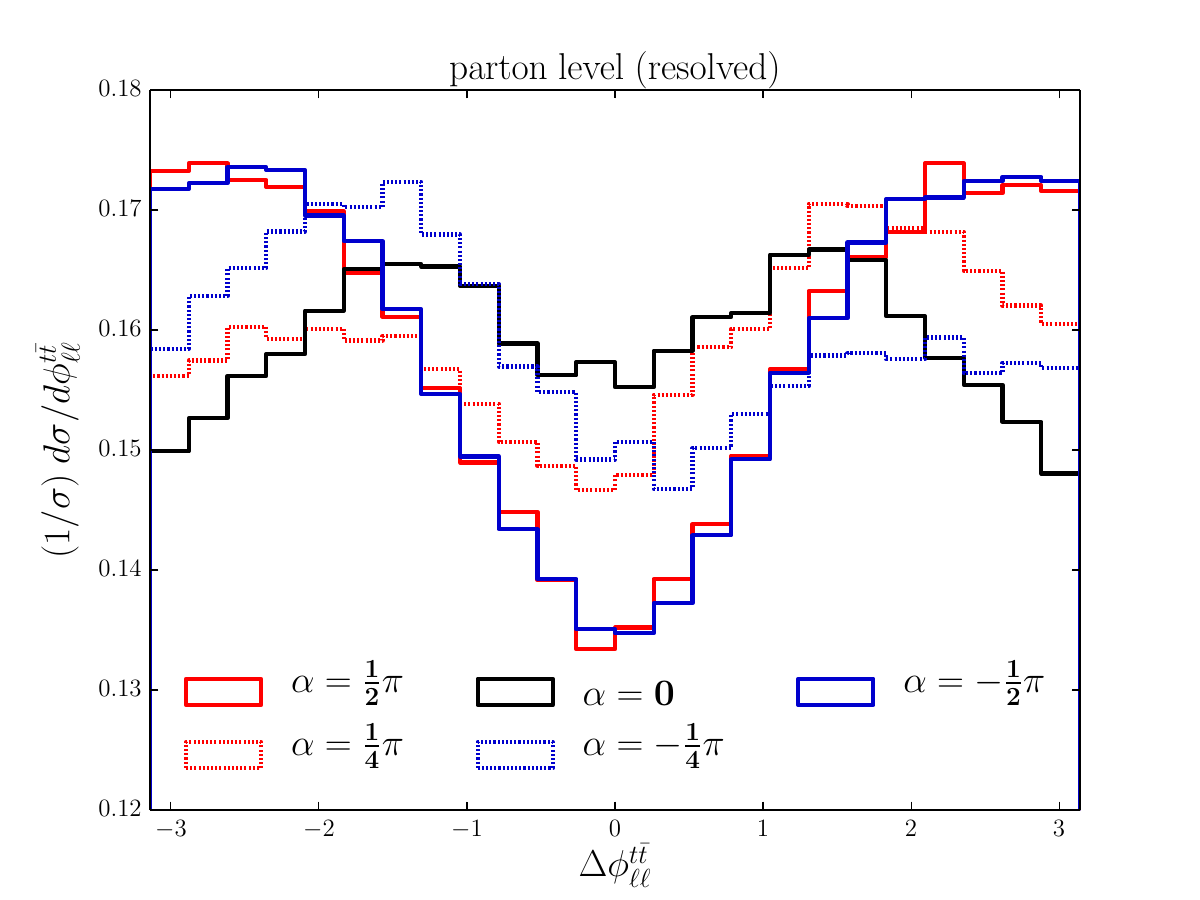} \hspace{-0.7cm} 
\includegraphics[scale=0.38]{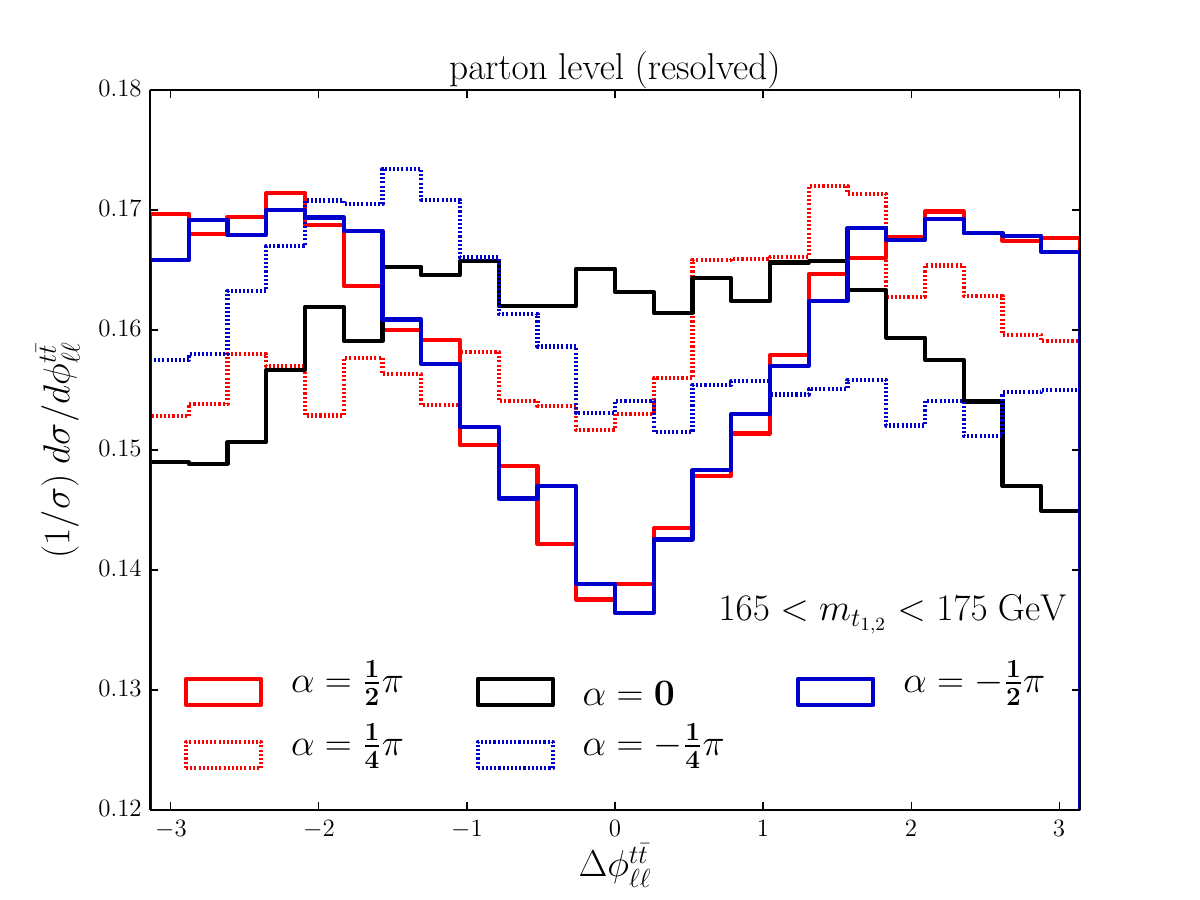} 
\caption{\label{fig:Delta_Phi_CM} \baselineskip 3.0ex 
$\Delta \phi^{t \bar{t}}_{\ell \ell }$ distributions for various choices of  CP-phase $\alpha$ at parton-level. The top pair rest-frame reconstruction is obtained via  the mass minimization procedure using  {\sc OPTIMASS}. No cuts are applied on the left plot, while the distributions in the right panel exploits a mass cut on $M_{2CW}^{(b\ell)}$ to improve the purity of the resolved events.} 
\end{figure}

In Fig.~\ref{fig:Delta_Phi_CM}, we present  $\Delta \phi^{t \bar{t}}_{\ell\ell }$ in the center-of-mass frame of the $t \bar{t}$ system (see Eq.~\ref{eq:obs2}) for various values of $\alpha$.
While Fig. \ref{fig:Delta_Phi_parton} assumes prior knowledge (parton-truth) of correct final state particles pairs,  Fig.~\ref{fig:Delta_Phi_CM} is obtained via the $M_2$ reconstruction. 
This distribution gets degraded as shown in the left panel of Fig.~\ref{fig:Delta_Phi_CM}, once we include all the resolved events (admixture of both correct and wrong combinations). 
However, one can make an improvement with a mass cut on $M_{2CW}^{(b\ell)}$ (see the bottom panel of Fig. \ref{fig:parton_momentum}.), ${165 \; {\rm GeV} < M_{2CW}^{(b\ell)} < 175 \; \rm{GeV} }$, and restore their original shapes, as shown in the right panel of Fig.~\ref{fig:Delta_Phi_CM}.

In the case of  CP mixed eigenstate ({\it e.g.} $ \alpha = \pm \pi/4$), the $\Delta \phi^{t \bar{t}}_{\ell  \ell } $ distributions are  asymmetric with respect to $\Delta \phi^{t \bar{t}}_{\ell \ell } = 0$. 
On the other hand, $\theta^\ast$ distributions are symmetric. 
Numerical values of $\Delta \phi^{t \bar{t}}_{\ell \ell }$ asymmetry are summarized in Table \ref{tab:Atable1_parton}. $\mathcal{A}_{\ell\ell} (\alpha=0,\pm\pi/2) = 0 $ is expected but we obtain nonzero values due to statistical uncertainties. We observe that the wrong combinatorics can be further suppressed with the $M_{2CW}^{(b\ell)}$ cut and  the resolved results become  closer to the idealistic parton-truth asymmetries. 

\begin{table}[t!]
\begin{center}
{\renewcommand{\arraystretch}{1.1}
\scalebox{1.0}{
\hskip -0.4cm \begin{tabular}{|c||c|c|c|}
\hline
         CP-phase                      & $\mathcal{A}_{\ell\ell}$ (parton-truth)  & $\mathcal{A}_{\ell\ell}$ (resolved)  & $\mathcal{A}_{\ell\ell}$ (resolved, cut)   \\ \hline\hline 
$ \alpha = \frac{1}{2} \pi $    &  $0.001 $                        &   $0.0005$                        & $0.0004$                        \\ 	\hline
$ \alpha = \frac{1}{4} \pi $    &  $0.032 $                          &   $0.021$                             & $0.027$                              \\ 	\hline
$ \alpha = 0 $                         &  $0.001$                        &   $-0.0002$                       & $-0.0005$                         \\ 	\hline
$ \alpha = -\frac{1}{4} \pi $    &  $-0.036$                        &   $-0.024$                            & $-0.031$                             \\ 	\hline
$ \alpha = -\frac{1}{2} \pi $    &  $-0.001$                      &   $-0.0008$                      & $-0.001$                           \\ 	\hline
\end{tabular}}}
\end{center}\vspace{-10pt}
\caption{Asymmetry variable $\mathcal{A}_{\ell\ell}$ for different CP phases calculated for three different samples at parton-level. 
Here ``resolved" samples include basic cuts only, while ``resolved, cut" samples include the mass cut $165 < m_{t_{1,2}} = M_{2CW}^{(b\ell)} < 175 \; \rm{GeV}$.
}
\label{tab:Atable1_parton}
\end{table}

   \subsection{Detector level analysis and LHC sensitivity~\label{sec:detector}}

After proving that our top mass reconstruction method dovetails nicely with CP-sensitive observables at the $t\bar{t}$ rest frame, we perform a full Monte Carlo study, including the Higgs boson decay to a pair of $b$-quarks. We require four bottom tagged jets and two opposite sign leptons in our signal. The major backgrounds for this signature in order of relevance  are $t \bar t b \bar b$ and $t \bar t Z$.

Both signal and SM backgrounds are simulated by the \texttt{MadGraph5\_aMC@NLO} with leading order accuracy in QCD at $\sqrt{s} = 14$ TeV. 
Higher order effects are included by normalizing the $tth$ rate to the  next-to-leading order (NLO)  QCD+EW cross-section 614~fb~\cite{deFlorian:2016spz},
and the  $t \bar t b \bar b$ and $t \bar t Z$ to their NLO QCD predictions 2.64~pb~\cite{Bredenstein:2009aj} and 1.06~pb~\cite{Maltoni:2015ena}, respectively. 
At generation level, we demand all partons to pass the following cuts:
\begin{eqnarray}
p_T > 20~{\rm GeV},\quad{\rm and}\quad ~| \eta |< 5~, \label{eq:basepartons}
\end{eqnarray}
 while leptons are required to have 
\begin{eqnarray}
p_T^\ell > 20~{\rm GeV},\quad{\rm and}\quad ~| \eta^\ell |< 2.5~. \label{eq:baseleptons}
\end{eqnarray}

Both signal and background events are showered and hadronized by \verb|PYTHIA 6| \cite{Sjostrand:2006za}. Jets are clustered with the \verb|FastJet| \cite{Cacciari:2011ma} implementation of the anti-$k_T$ algorithm~\cite{Cacciari:2008gp} with a fixed cone size of $R = 0.4 \;(1.2)$ for a slim (fat) jet. We include simple detector effects based on the ATLAS detector performances~\cite{ATL-PHYS-PUB-2013-004}, and smear momenta and energies of reconstructed jets and leptons according to their energy values. See Appendix~\ref{app:DR} for more details.

In the phase space where the Higgs is kinematically boosted, its decay products are collimated in the same direction. In this regime, the Higgs can be better reconstructed using a single fat jet evading its possible intervention to the $t \bar{t}$-system. Therefore, our previous method of resolving a combinatorial problem can be  repeatedly applicable in the boosted Higgs configuration. 

The boosted Higgs jet with a two-pronged substructure is a rare feature that the SM backgrounds retain. Thus, it delivers a further handle to disentangle the backgrounds from our signal events. The first demonstration of the use of a jet substructure technique in the dileptonic  $t \bar t h (b \bar{b})$ channel can be found in Ref.~\cite{Buckley:2015vsa}, where it effectively kills both $t \bar t b \bar b$ and $t \bar t Z$ backgrounds. Here we follow similar steps, employing the \verb| TemplateTagger v.1.0 |~\cite{Backovic:2012jk} implementation of the Template Overlap Method (TOM)~\cite{Almeida:2010pa,  Backovic:2013bga} as a boosted Higgs tagger, due to its robustness against pile-up contaminations.

We first require at least one $R= 1.2$ fat jet with 
\begin{eqnarray}
p_T^J > 200 \gev,\quad{\rm and}\quad |\eta^J| < 2.5.\label{eq:basefatjet}
\end{eqnarray}
For a fat jet to be tagged as a Higgs, we demand a two-pronged Higgs template overlap score 
\begin{eqnarray}
Ov_2^h > 0.5.\label{eq:fatH}
\end{eqnarray}
We require exactly one Higgs-tagged fat jet that passes the cuts in Eqs.~(\ref{eq:basefatjet}-\ref{eq:fatH}) and has $2b$-tagged slim jets inside~\footnote{\label{fn:btag} In our $b$-tagging algorithm, $R=0.4$ jets are classified into three categories: If a $b$-hadron ($c$-hadron) is found inside a slim jet, it is classified as a $b$-jet ($c$-jet). The remaining unmatched jets are called light-jets. Each jet candidate is multiplied by an appropriate tag-rate~\cite{ATL-PHYS-PUB-2016-026}. We apply a flat $b$-tag rate of $\epsilon_{b \rightarrow b} = 0.7$ and a mis-tag rate that a $c$-jet (light-jet) is misidentified as a $b$-jet of $\epsilon_{c \rightarrow b} = 0.2$ ($\epsilon_{j \rightarrow b} = 0.01)$. For a $R= 1.2$ fat jet to be $b$-tagged, we require that a $b$-tagged slim jet is found inside a fat jet. To take into account the case where more than one $b$-jet lands inside a fat jet, we reweight a $b$-tagging efficiency based on a scheme described in Ref.~\cite{Backovic:2015bca}.}:
\begin{eqnarray}
N_h =1.\label{eq:NTH}
\end{eqnarray}
Additionally, we require at least two slim jets that are isolated from the Higgs-tagged fat jet 
\begin{eqnarray}
p_T^j > 30 \gev,\quad{\rm and}\quad|\eta^j| < 2.5,\label{eq:basethinjet}
\end{eqnarray}
in which we require exactly two $b$-tagged slim jets. We demand exactly two isolated leptons passing the cuts in Eq.~(\ref{eq:baseleptons}) and
\begin{eqnarray}
 p_T^\ell / p^{\Sigma}_T > 0.7,\label{eq:isolepton}
\end{eqnarray}
where $p^{\Sigma}_T$ is the sum of transverse momenta of final state particles (including a lepton) within $\Delta R = 0.3$ isolation cone.
\begin{figure}[t!]
\centering
\includegraphics[scale=0.38]{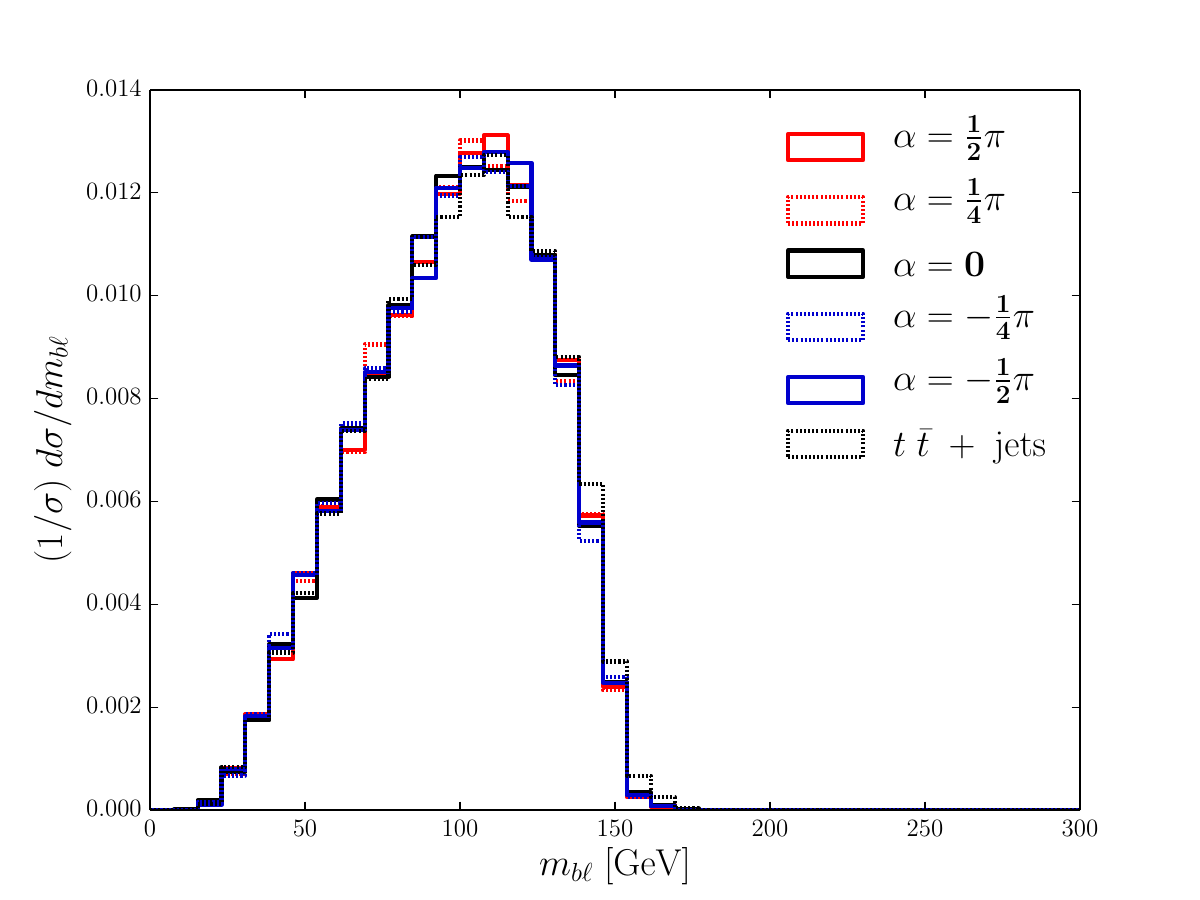}\hspace{-0.6cm}
\includegraphics[scale=0.38]{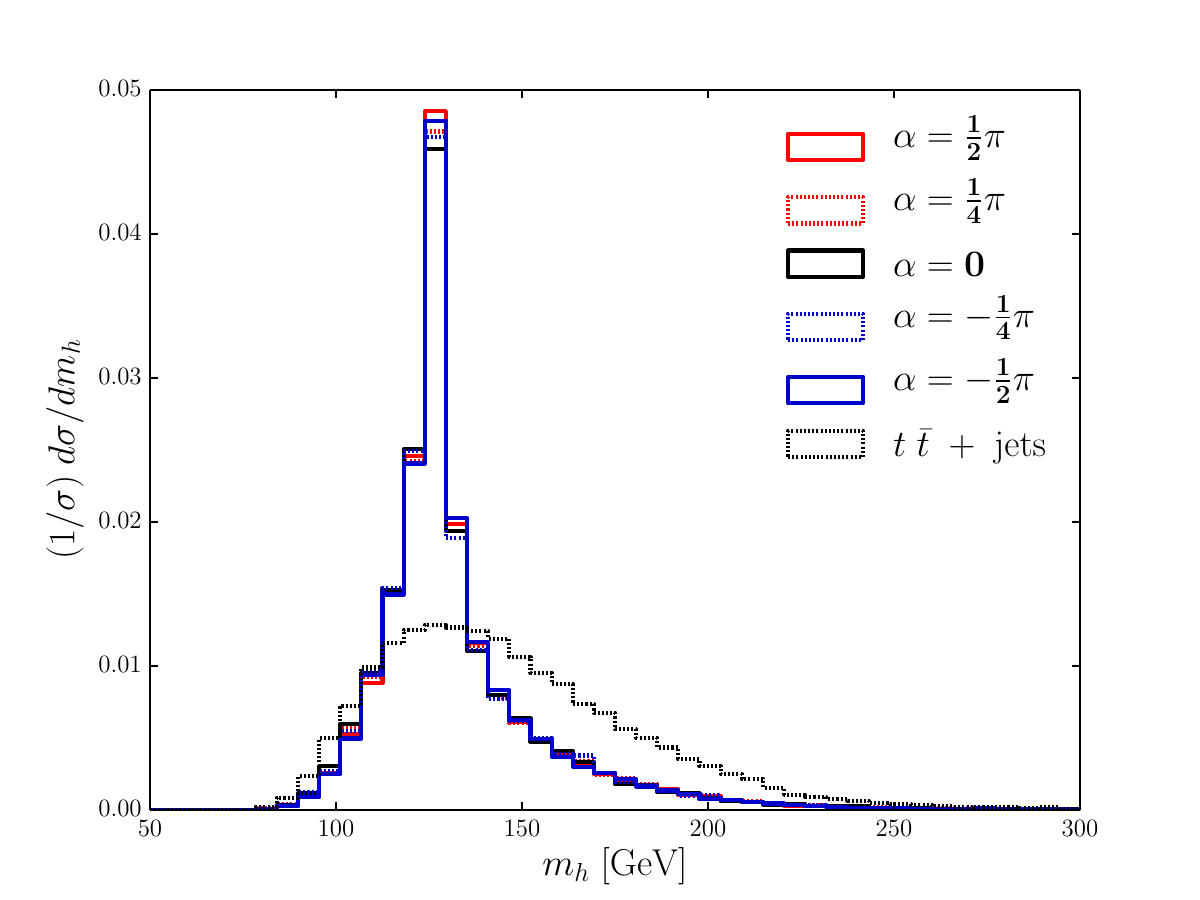} \\
\includegraphics[scale=0.38]{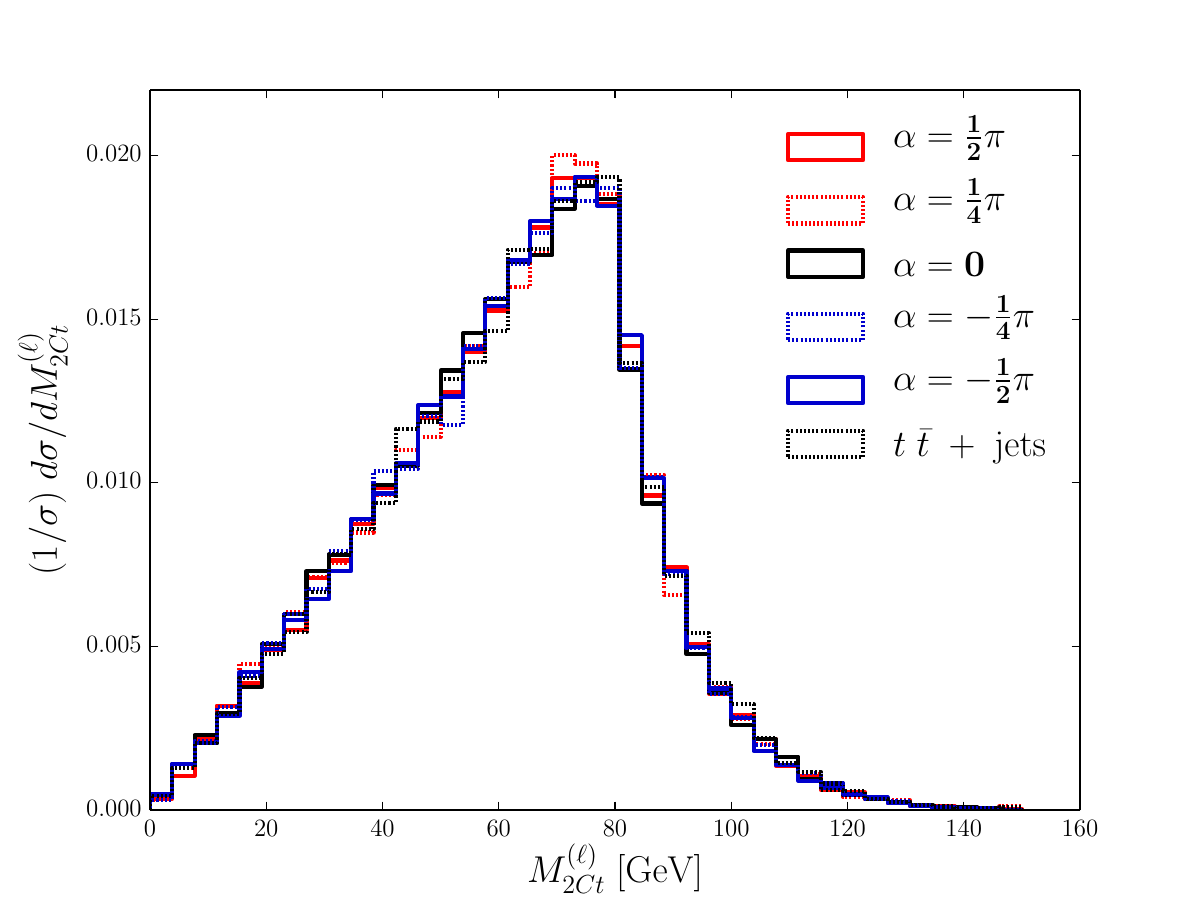}\hspace{-0.6cm}
\includegraphics[scale=0.38]{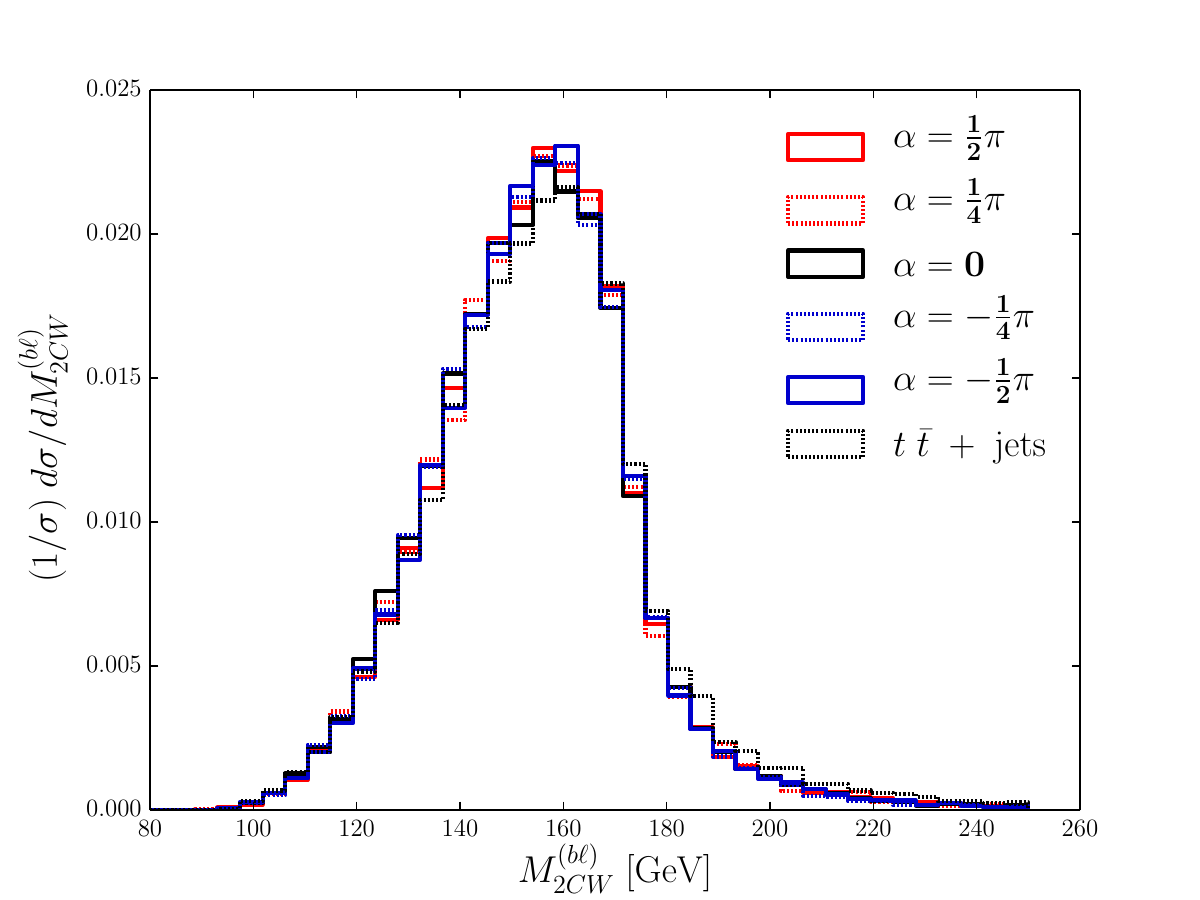} 
\caption{\label{fig:dist_detector} \baselineskip 3.0ex 
Higgs tagged fat jet reconstructed mass $m_{h}$ (upper-left) and $m_{b\ell}$ (upper-right)  distributions after the boosted selection for different CP phases. We also show fully reconstructed $M_{2Ct}^{(\ell)}$ (lower-left)  and $M_{2CW}^{(b\ell)}$ (lower-right)  distributions. All plots are generated after resolving the combinatorial problem and 4$b$-tagging. 
} 
\end{figure}

In Fig.~\ref{fig:dist_detector}  (upper-left), we show the reconstructed invariant mass distributions for Higgs-tagged fat jet, laid out with the dominant $t \bar t b \bar b$ background. The distributions are insensitive to different CP structures, but provide more separation from the background. Hence, we select the Higgs mass window
\begin{eqnarray}
105~\gev< m_h < 145~\gev \;. \label{eq:mhwindown}
\end{eqnarray}   
The other reconstructed invariant mass distributions   $m_{b\ell}$ (upper-right), $M_{2Ct}^{(\ell)}$ (lower-left)  and $M_{2CW}^{(b\ell)}$ (lower-right)  are also shown in Fig.~\ref{fig:dist_detector}. The distribution of reconstructed $M_{2CW}^{(b\ell)}$ becomes broader due to parton shower, hadronization and detector resolution effects, compared to parton-level results in Fig. \ref{fig:dists}, but the basic shape remains the same. 
\begin{figure}[t!]
\centering
\includegraphics[scale=0.38]{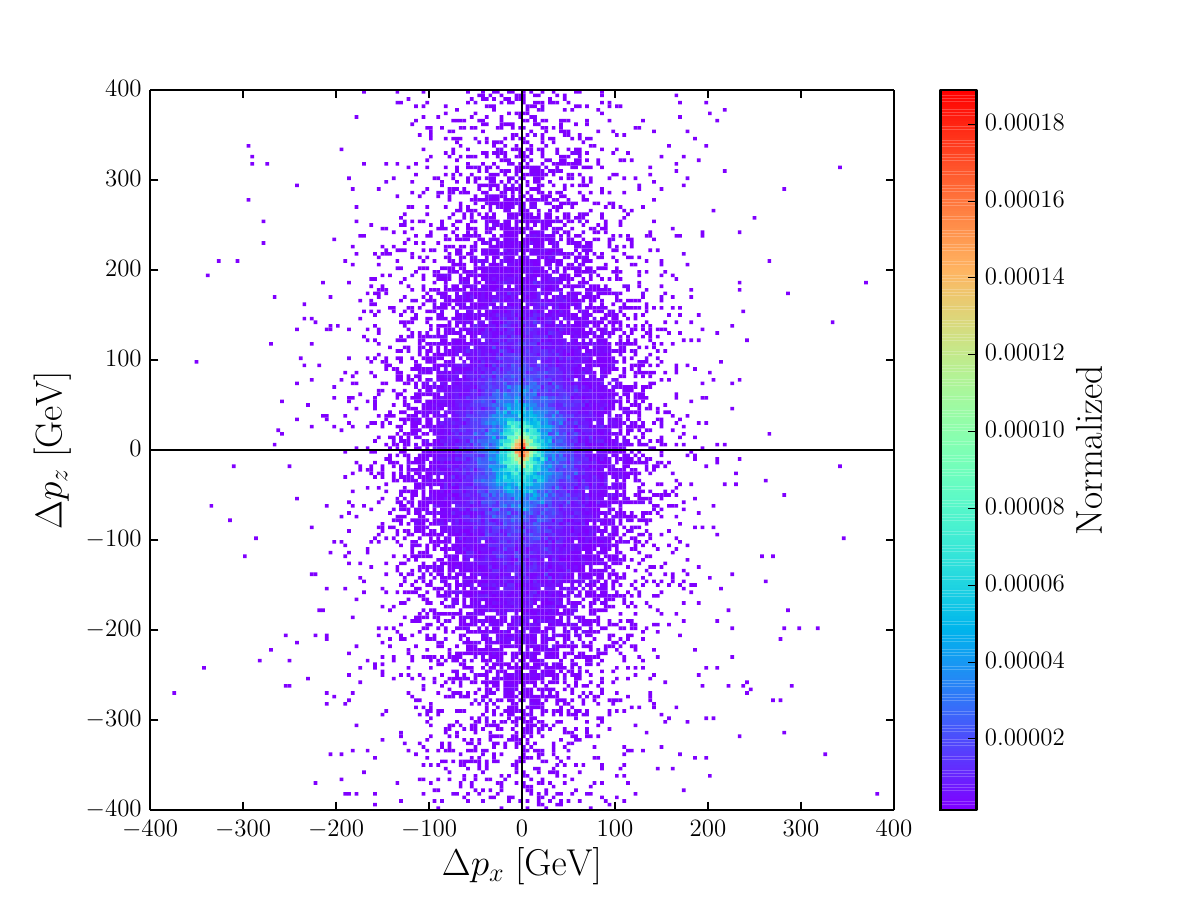} \hspace{-0.6cm}
\includegraphics[scale=0.38]{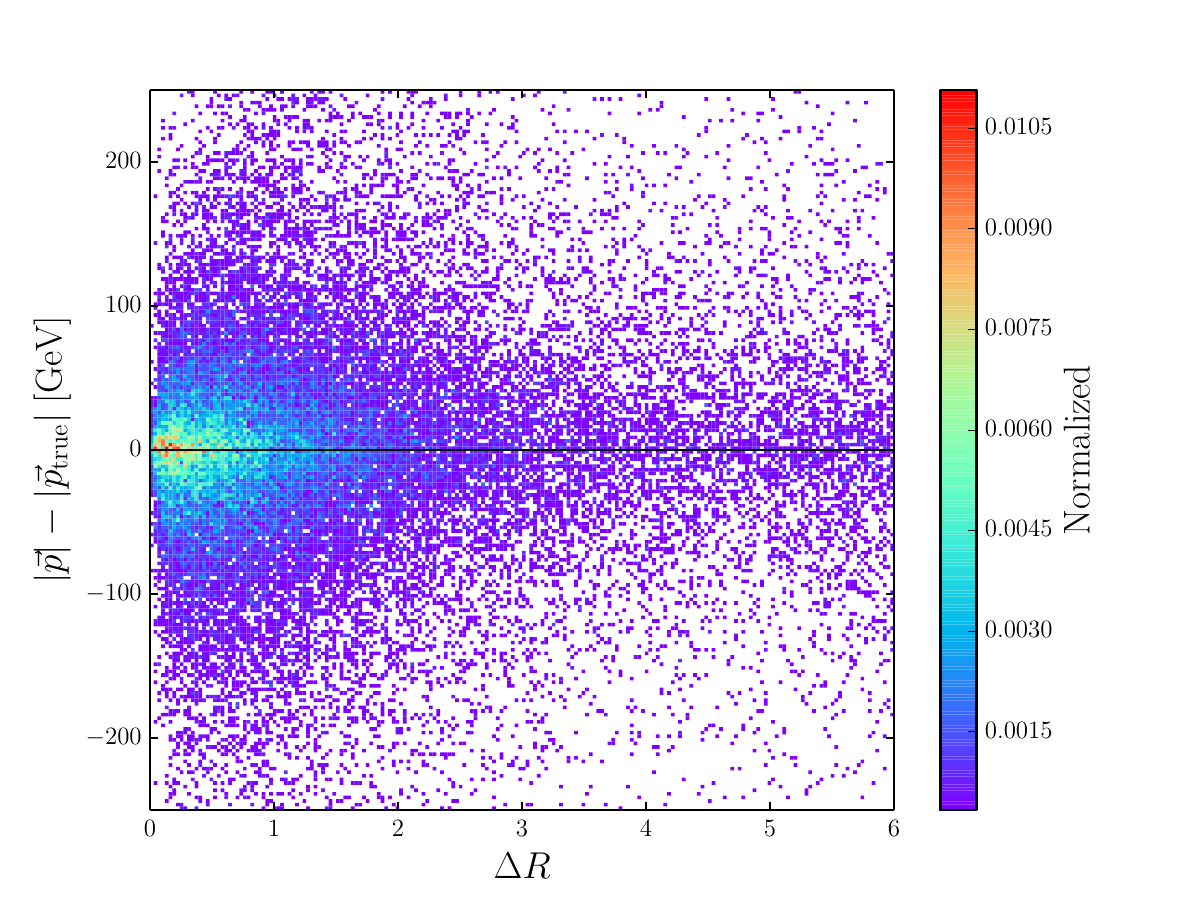}  \\
\includegraphics[scale=0.38]{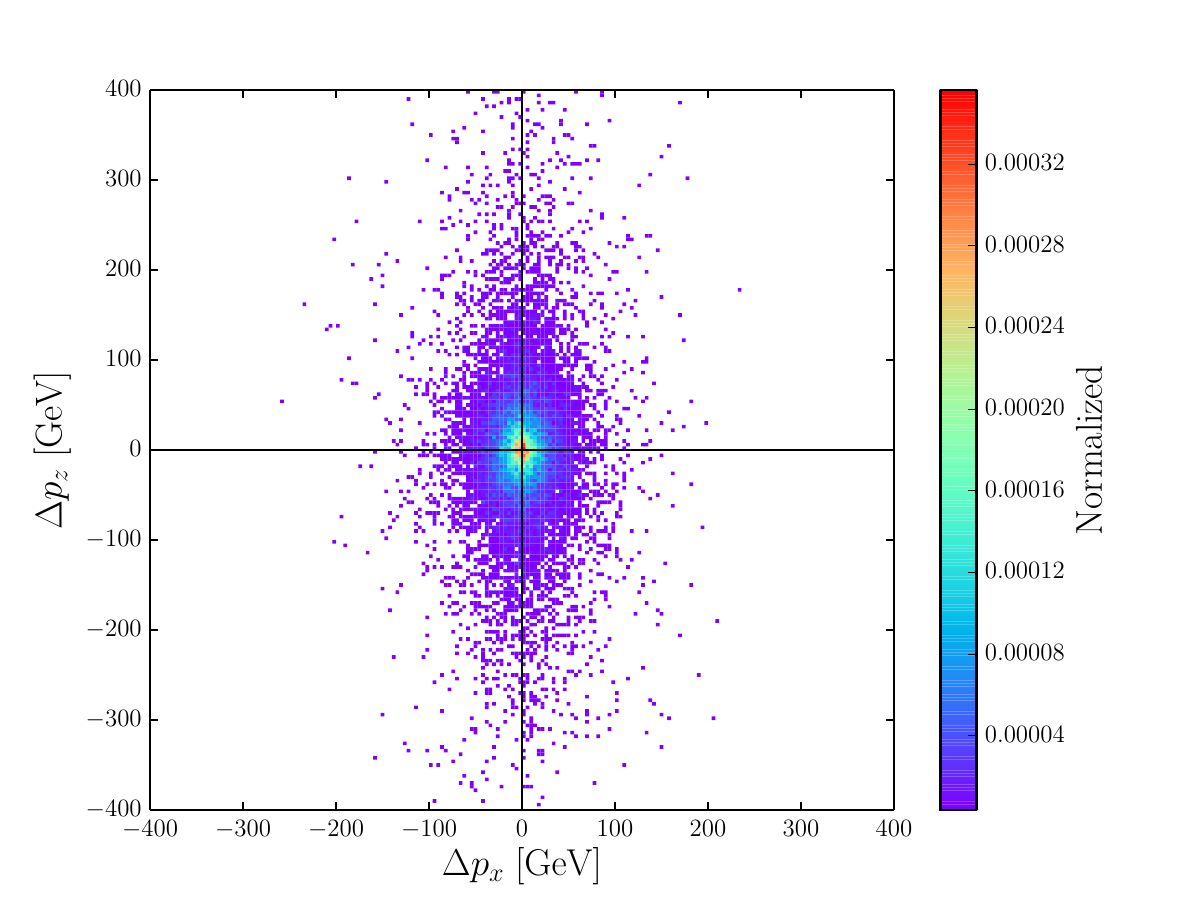} \hspace{-0.6cm}
\includegraphics[scale=0.38]{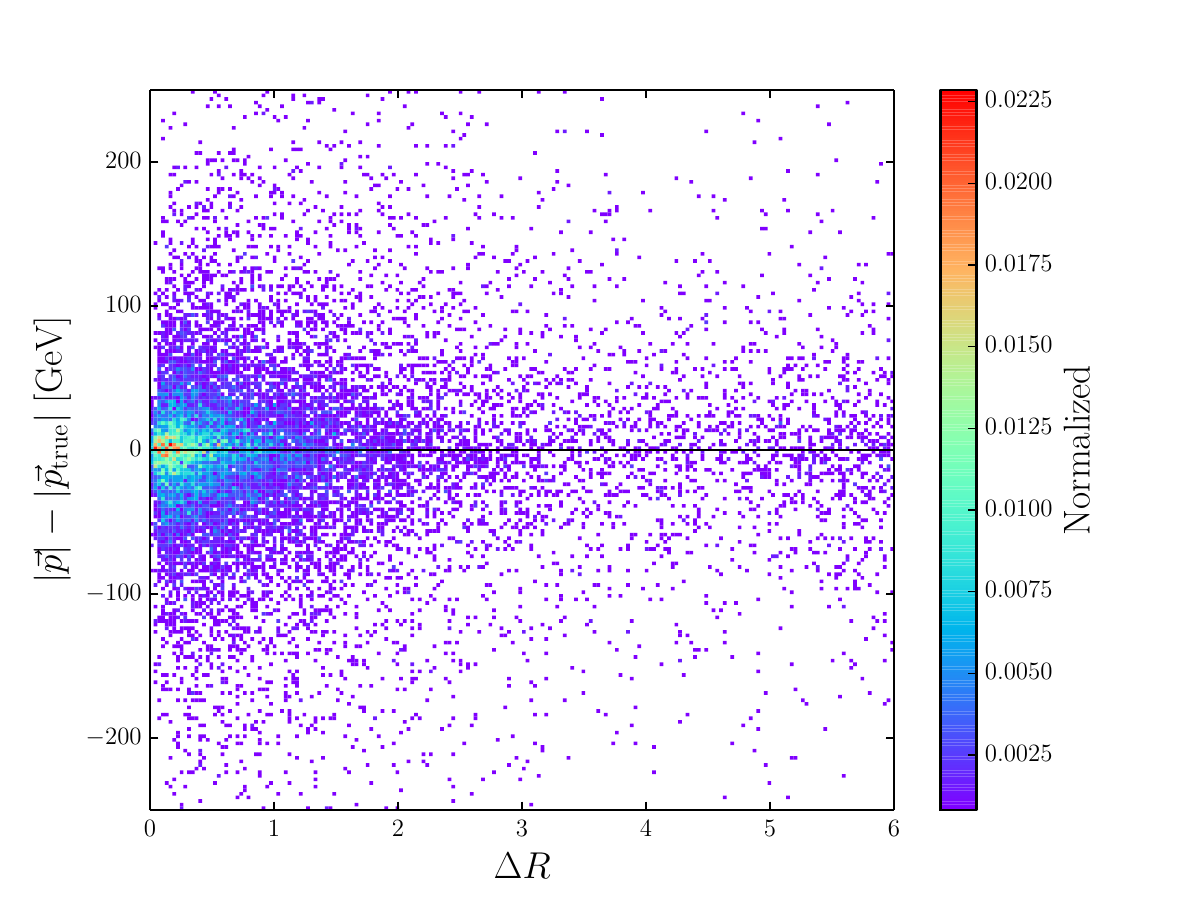}  
\caption{\label{fig:detector_momentum} \baselineskip 3.0ex 
Correlations between (left panels) $\Delta p_z$ and $\Delta p_x$ and (right panels) $| \vec p \; | - | \vec p_{true} |$ and $\Delta R( \vec p ,  \vec p_{true})$ for $M_{2CW}^{(b\ell)}$ with respect to $\alpha = 0$ case. All plots are generated after resolving the combinatorial problem, 4$b$-tagging and $105~\gev < m_h < 145~\gev$, without (top panels)  and  with (bottom panels) an additional mass cut $155 ~{\rm GeV} < M_{2CW}^{(b\ell)} < 180 ~{\rm GeV}$.
} 
\end{figure}

We resolve the combinatorial ambiguity of the two $b$-lepton pairs based on the prescription in Eq.~(\ref{setofthreeWt}). The efficiency of the method for our signal is $82\%$ (comparable to the efficiency at parton level), yet at the same time $t \bar t b \bar b$ and $t \bar t Z$ backgrounds are cut down to $64\%$ and $70\%$, respectively. Hence, the top mass reconstruction method works as an extra relevant handle in the background suppression, eliminating  wrong combinations from  b-jets that are not from the top decays.

Momentum reconstructions of two neutrinos are displayed in Fig.~\ref{fig:detector_momentum}. The level of accuracy in reconstructing neutrino momenta also degrades to some extent, where the uncertainty in $p_z$ direction is greater than the transverse components.
Additional mass cut ${155~\gev < M_{2CW}^{(b\ell)} < 180}$~GeV reduces the reconstruction efficiency to $\epsilon = 32\%$, but would increase the purity of the sample and improve the momentum resolution. We observe that the reconstruction method is robust to parton-shower, hadronization, and detector resolution effects, presenting similar efficiencies to the parton level analysis. Our reconstruction is better than (or comparable to) existing results. For example, Ref.~\cite{AmorDosSantos:2017ayi} performs a conventional kinematic mass reconstruction with the missing transverse momentum and attempts resolving the two-fold sign ambiguity using a likelihood based on transverse momenta of the involved particles. This method leads to 62\% efficiency with 50\% purity for signal, and 51\% efficiency for backgrounds. Since our method is purely based on mass minimization, it is less sensitive to new physics modifications and is a suitable element for a robust spin-correlation analysis. We note that one can further improve the efficiency of our method by utilizing those discarded ``unresolved'' events and deploying a hybrid method \cite{Debnath:2017ktz} together with $M_2$ reconstruction. 

We acknowledge that there is a certain degree of uncertainty in the precision compared to parton-level results in Fig. \ref{fig:parton_momentum}, where the peaks are broadened. We attribute this change to contaminations in the total missing transverse momentum where additional neutrinos from $h\rightarrow b \bar{b}$ system, via the semi-leptonic decays of the $b$-hadrons, can disrupt the relations in Eqs.~(\ref{eq:m2CWdef})-(\ref{eq:m2Ctdef}), in combination with detector effects. Nevertheless, overall net shapes stay the same showing its resilience over the procedures.

\begin{figure}[!t]
\includegraphics[scale=0.38]{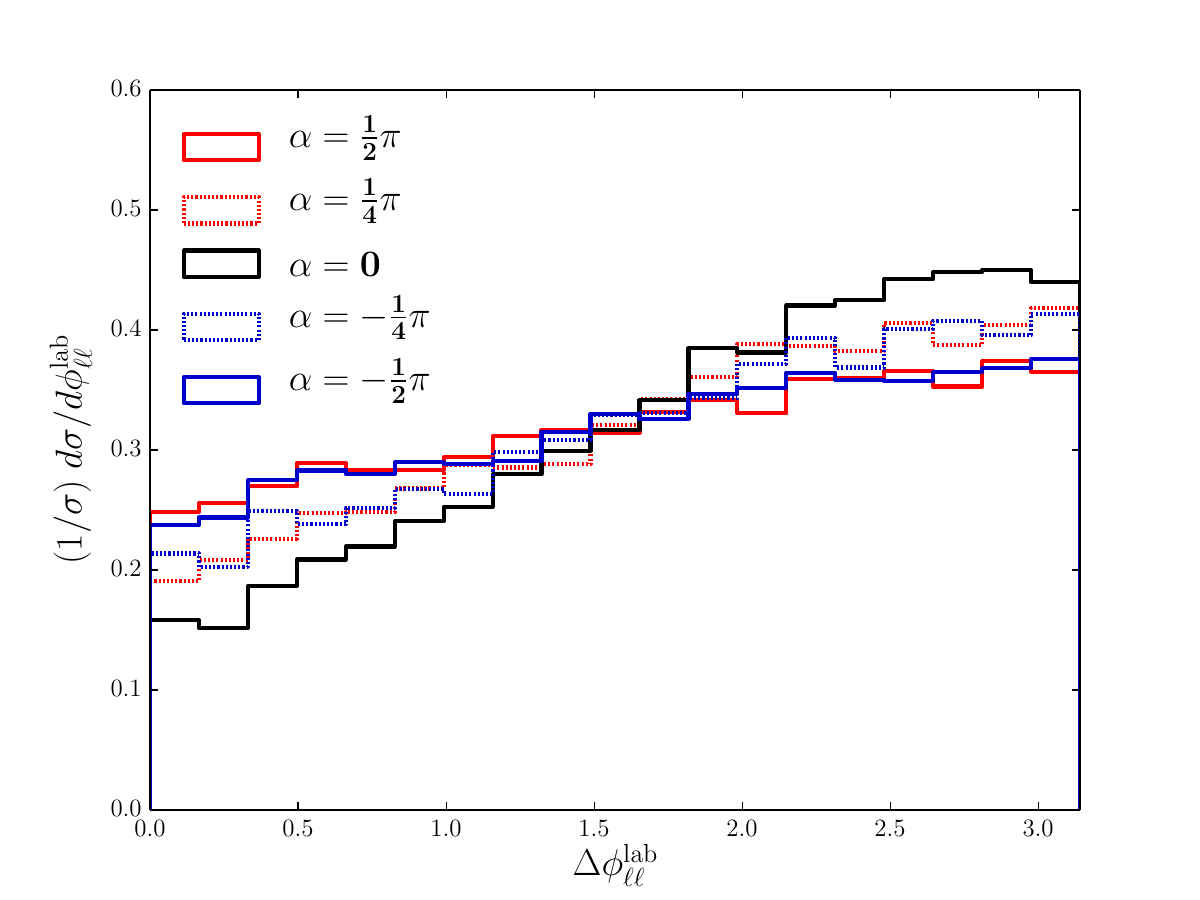}
\includegraphics[scale=0.38]{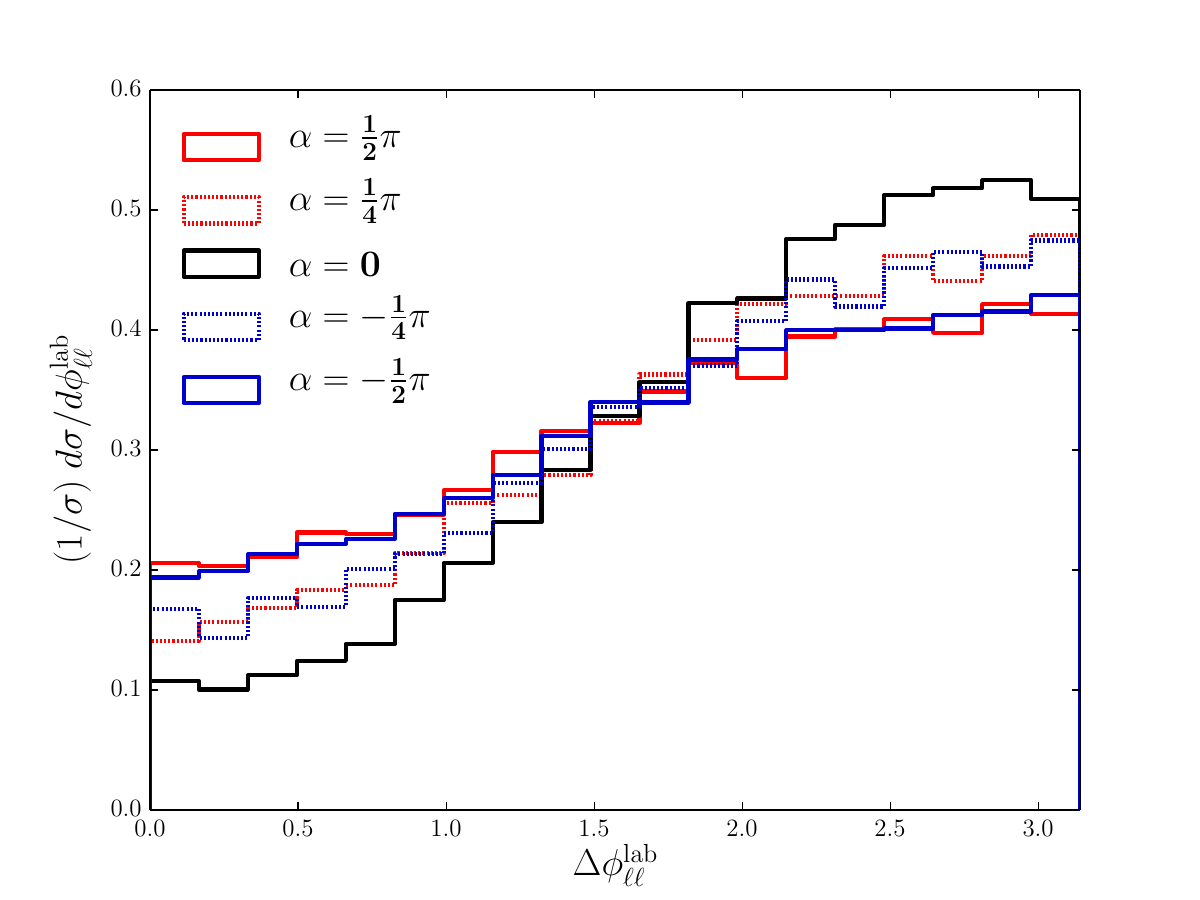} 
\caption{\label{fig:deltaphi_lab_detector} \baselineskip 3.0ex 
Distributions of $\Delta \phi^{\rm lab}_{\ell \ell}$, after resolving the combinatorial problem and 4$b$-tagging, without (left)  and  with (right) an additional $m_{\ell\ell} > 75 ~{\rm GeV}$ selection.
} 
\end{figure}
\begin{figure}[!t]
\includegraphics[scale=0.38]{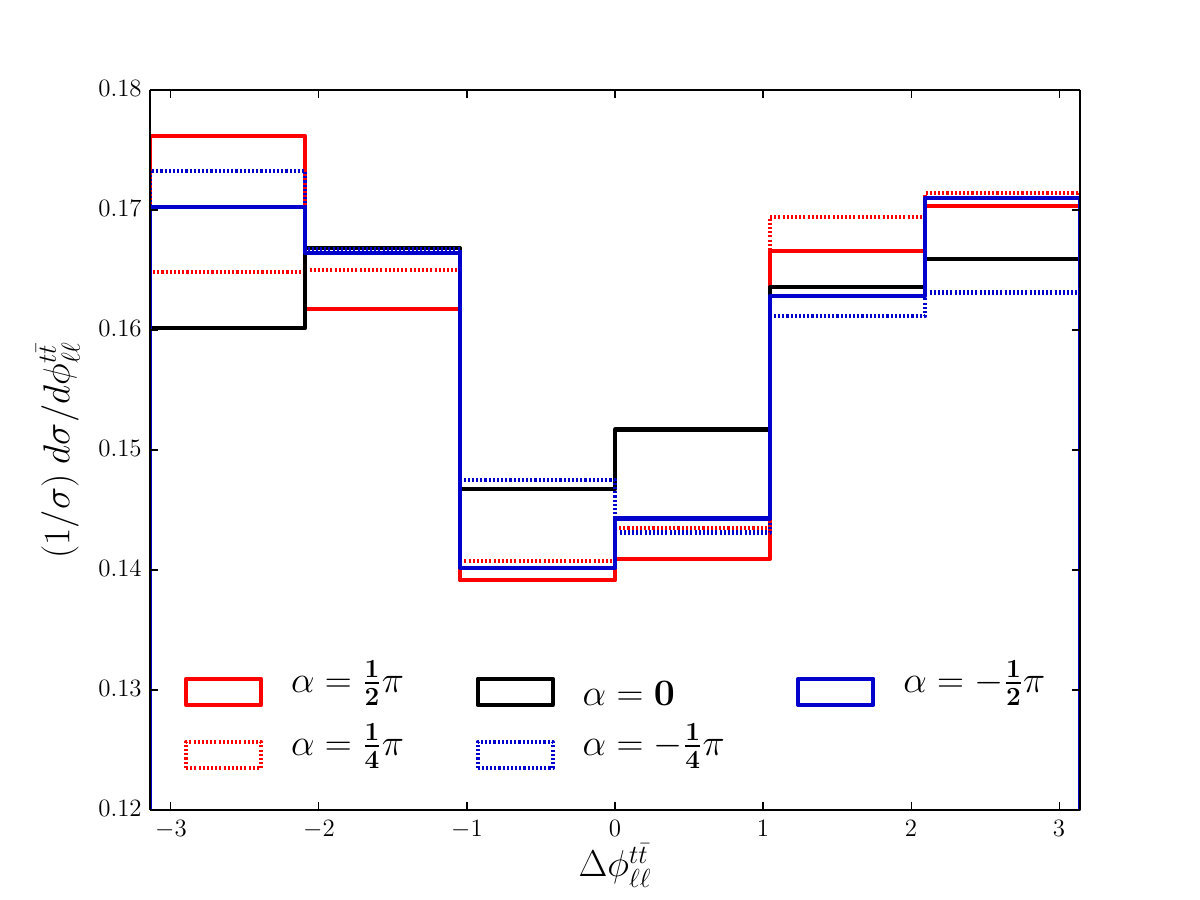} \hspace{-0.6cm}
\includegraphics[scale=0.38]{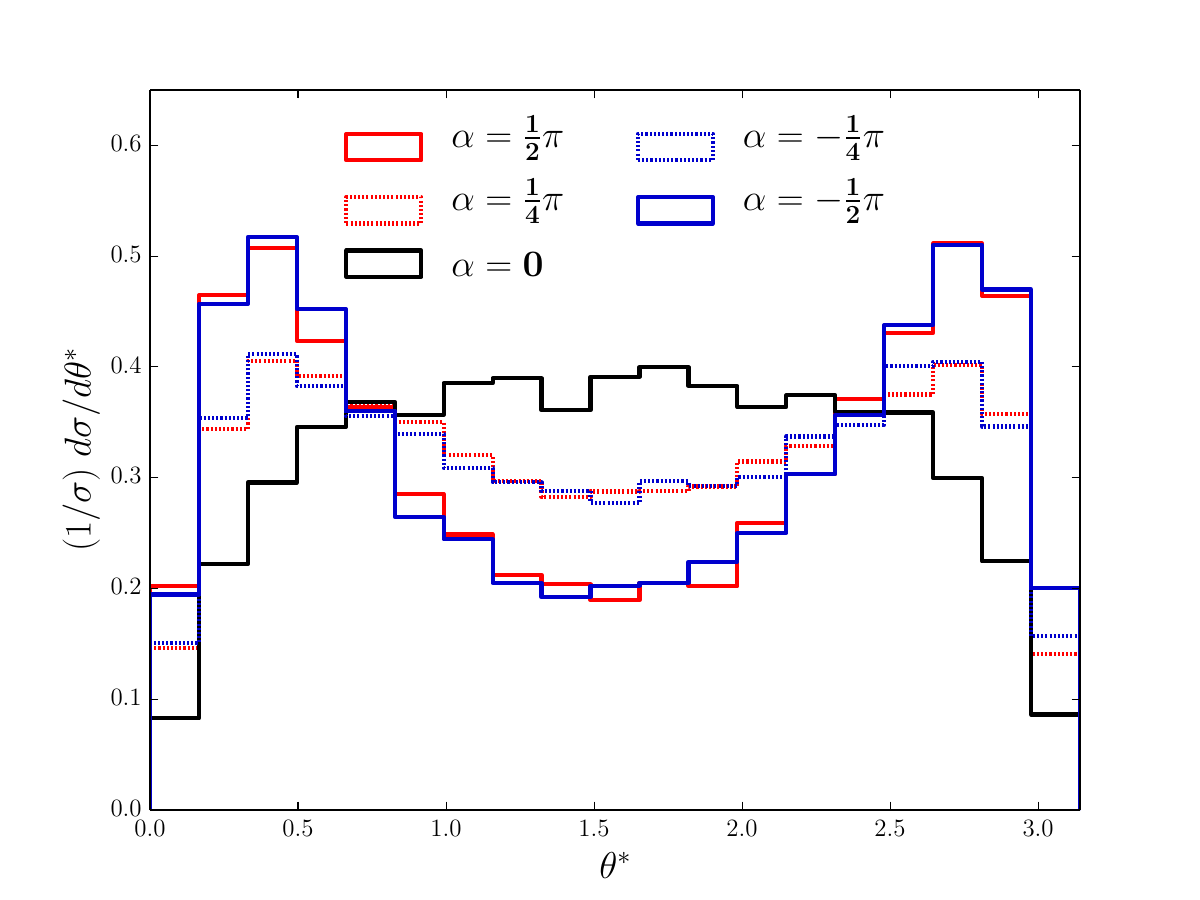} \\
\includegraphics[scale=0.38]{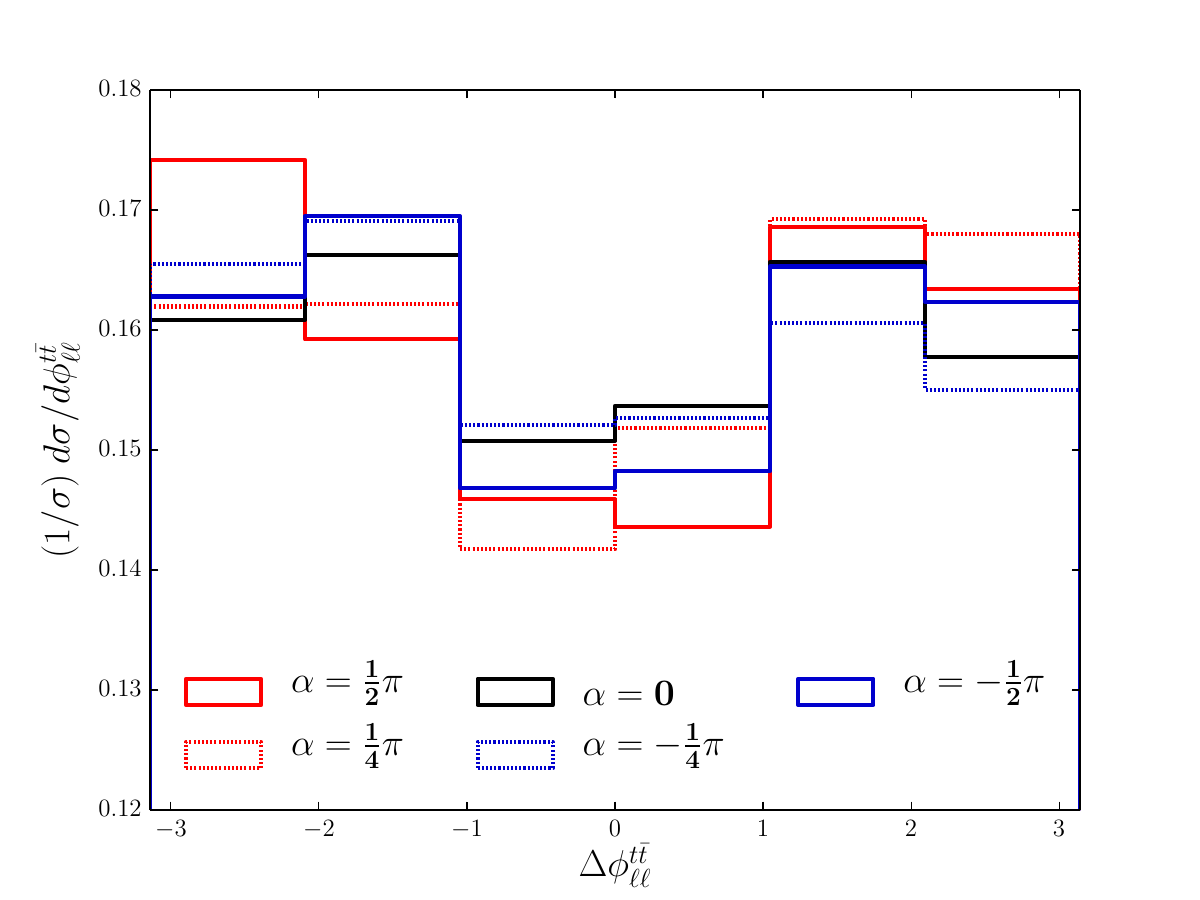} \hspace{-0.6cm}
\includegraphics[scale=0.38]{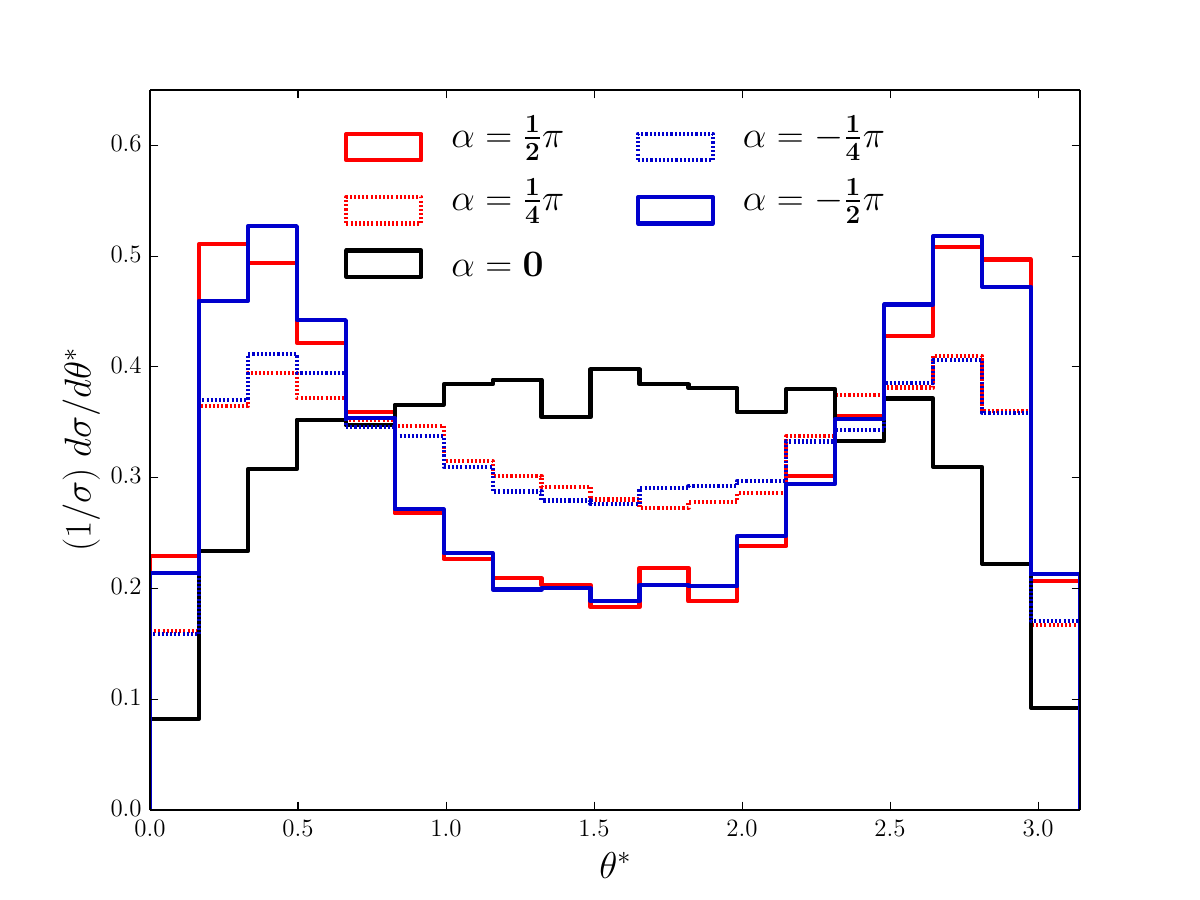} 
\caption{\label{fig:cosdetector} \baselineskip 3.0ex 
$\Delta \phi^{t \bar{t}}_{\ell \ell }$ (left panels)  and  $\theta^\ast$ (right panels) distributions, after resolving the combinatorial problem, 4$b$-tagging and $105~\gev< m_h < 145~\gev$,  without (top panels) and with (bottom panels) an additional mass cut $155 ~{\rm GeV} < M_{2CW}^{(b\ell)} < 180 ~{\rm GeV}$. 
} 
\end{figure}
Distributions of $\Delta \phi^{\rm lab}_{\ell \ell }$, $\Delta \phi^{t \bar{t}}_{\ell \ell }$, and $\theta^\ast$ are presented in Figs. \ref{fig:deltaphi_lab_detector} and \ref{fig:cosdetector}. 
The $\theta^\ast$ and $\Delta \phi^{\rm lab}_{\ell \ell }$ distributions remain very similar to those at parton level (Fig. \ref{fig:Delta_Phi_parton} and Fig. \ref{fig:cosparton}), while $\Delta \phi^{t \bar{t}}_{\ell \ell }$ distribution gets more distorted (see Fig. \ref{fig:Delta_Phi_CM}).

Table~\ref{tab:Cutflow1} summarizes the impact of a series of cuts for the signal ($\alpha = 0$) and background cross sections. In the last column, we show the significances ($\sigma$), which are calculated for a luminosity of 3~$\rm{ab}^{-1}$, using the expression 
\beq
  \sigma \equiv
    \sqrt{-2\,\ln\bigg(\frac{L(B | S\!+\!B)}{L( S\!+\!B| S\!+\!B)}\bigg)}\,,
  \ \text{with}\
  L(x |n) =  \frac{x^{n}}{n !} e^{-x} \,,
\label{Eq:sigfinicance} \eeq
where $S$ and $B$ are the expected number of signal and background events, respectively~\cite{Cowan:2010js}. 
We find that our results are roughly in agreement with those from Ref. \cite{Buckley:2015vsa}. 
Although we obtain high significance as shown in the first row of Table~\ref{tab:Cutflow1}, we would impose more stringent cuts for high-purity sample of $t\bar t h$ production. We obtain $\sigma=8.1$ with the resolved combinatorics. For an additional mass cut, we could retrieve even higher purity but we would suffer from statistics. In the following analysis, we do not impose this mass cut but instead require the dilepton invariant mass cut, $m_{\ell\ell} >75$ GeV. 
The asymmetry results at detector-level are summarized in Table~\ref{tab:Atable1_detector}. They can be compared against those at parton-level in Tables \ref{tab:Atable1_parton}. 

\begin{table*}[t]
\begin{center}
\setlength{\tabcolsep}{0.9mm}
\renewcommand{\arraystretch}{1.3}
\scalebox{1.0}{
\hspace*{-20pt}
\begin{tabular}{|c||c|c|c|c|c|}
\hline
    cuts                                                                                                     & $t \bar{t} h $ ($\alpha = 0$)   & $t \bar{t} b \bar{b} $           & $t \bar{t} Z $         & $S/B$        & $\sigma$        \\  \hline \hline
$N_{h}= 1$, $4b$-tags, $p_T^\ell  > 20 \gev$, $|\eta^\ell | < 2.5$  &  \multirow{2}{*}{ $0.075$ }      &   \multirow{2}{*}{ 0.25 }         &  \multirow{2}{*}{0.012}  &  \multirow{2}{*}{ 0.23 }   &  \multirow{2}{*}{ 6.64 }       \\   
$p_T^j  > 30 \gev$, $|\eta^j| < 2.5$, $N_{j} \geq 2$, $N_{\ell} = 2$    &                                               &                                            &                                    &                  &                 \\  \hline
$105~\gev < m_h < 145 \gev$                                                                       & $0.056$                               & $0.12$                                 & $0.0067$              &   0.35   &     $7.00$  \\  \hline
Resolving combinatorics                                                                       & $0.046$                               & $0.077$                               & $0.0047$                & 0.45    &    $7.07$    \\  \hline 
$m_{\ell \ell} > 75 ~{\rm GeV}$                                                              & $0.038$                               & $0.058$                               & $0.0038$                 &  0.49  &    $6.68$    \\  \hline 
\end{tabular}}
\end{center}
\caption{Cumulative cut-flow table showing the SM background and signal ($\alpha = 0$) cross sections in fb. Significances ($\sigma$) are calculated for a luminosity of 3~$\rm{ab}^{-1}$.}
\label{tab:Cutflow1}
\end{table*}

\begin{table}[h]
\begin{center}
{\renewcommand{\arraystretch}{1.1}
\scalebox{1.0}{
\hskip -0.4cm \begin{tabular}{|c||c|c|}
\hline
 CP-phase                              &  $\mathcal{A}_{\ell\ell}$    & $\mathcal{A}_{\ell\ell}$ (cut)     \\ \hline \hline
$ \alpha = \frac{1}{2} \pi $    &   $0.001$           & $-0.004$                                     \\ 	\hline

$ \alpha = \frac{1}{4} \pi $    &  $0.015$               & $0.024$                                               \\ 	\hline

$ \alpha = 0 $                       &   $0.007$           & $-0.001$                                       \\ 	\hline

$ \alpha = -\frac{1}{4} \pi $    &   $-0.021$            & $-0.020$                                             \\ 	\hline

$ \alpha = -\frac{1}{2} \pi $    &  $0.001$           & $-0.003$                                         \\ 	\hline 

\end{tabular}}}
\end{center}\vspace{-10pt}
\caption{Asymmetry variables $\mathcal{A}_{\ell\ell}$ after resolving the combinatorial problem, 4$b$-tagging and ${105\gev < m_h < 145 \gev}$, without and with an additional mass cut ${155 ~{\rm GeV} < M_{2CW}^{(b\ell)} < 180 ~{\rm GeV}}$. 
}
\label{tab:Atable1_detector}
\end{table}

\begin{figure}[!t]
\centering
\includegraphics[scale=0.4]{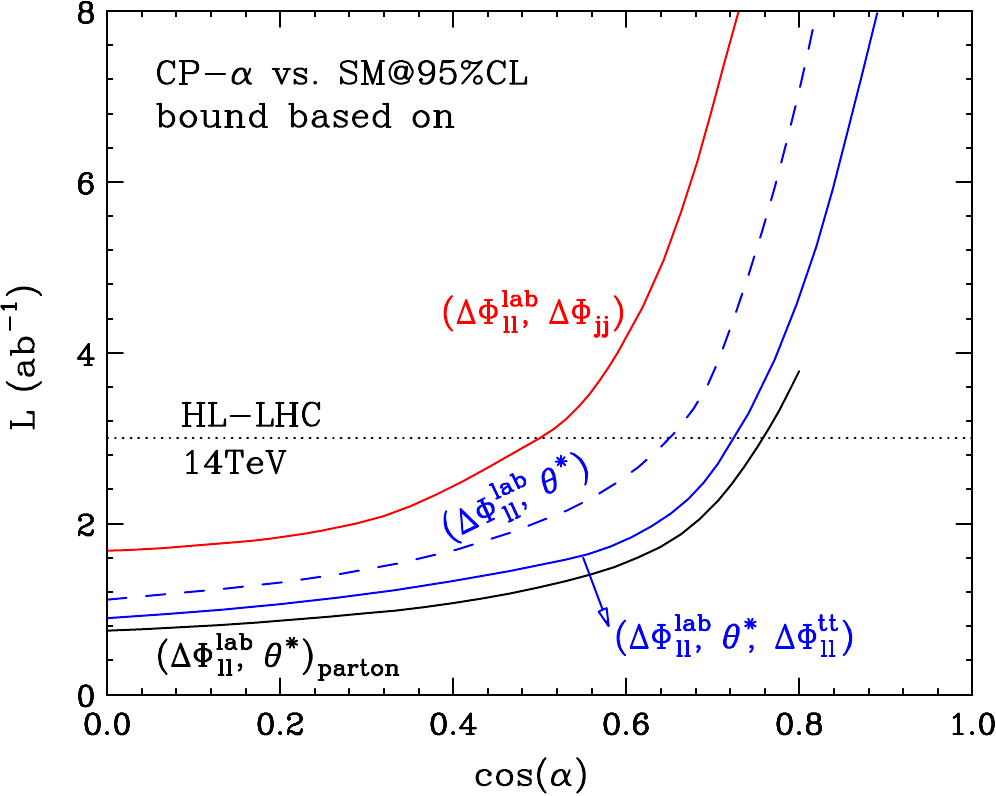} \hspace{0.5cm}
\includegraphics[scale=0.5]{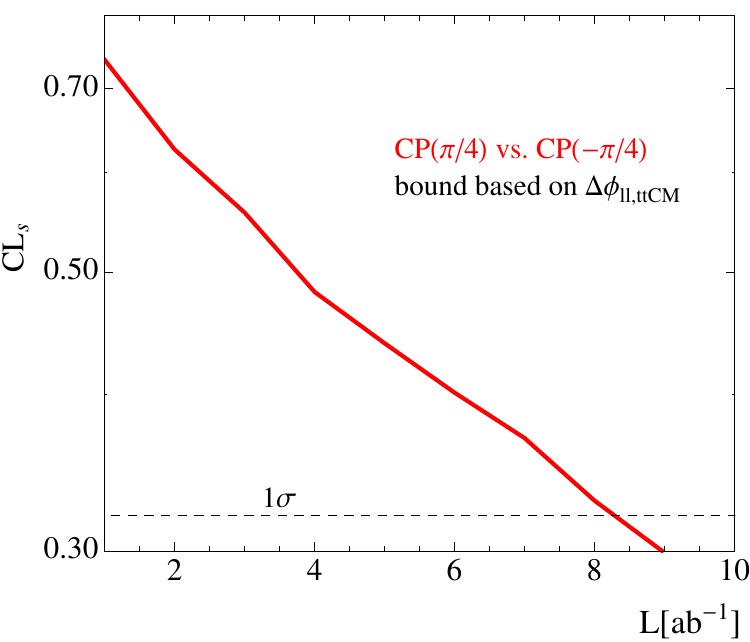} 
\caption{\label{fig:tthCP_bound} \baselineskip 3.0ex 
Left: Luminosity required to distinguish an arbitrary CP-$\alpha$ state from the SM Higgs via $t\bar{t}h$ production. 
Our limits are based on a binned log-likelihood analysis for $(\Delta\phi^{\rm lab}_{\ell\ell},\Delta\phi_{jj})$ (red full), 
$(\Delta\phi^{\rm lab}_{\ell\ell},\theta^*)$ (blue dashed), and $(\Delta\phi^{\rm lab}_{\ell\ell},\theta^*, \Delta\phi^{t\bar t}_{\ell\ell})$ (blue full),
accounting for the full detector level analysis. To illustrate the robustness of the top reconstruction method when going from the parton to the 
detector level, we also show the bounds using the parton-level distributions $(\Delta\phi^{\rm lab}_{\ell\ell},\theta^*)$  with the rates rescaled to the full detector analysis (black full). Right: CL$_s$ as a function of the luminosity to distinguish CP$(\pi/4)$ from CP$(-\pi/4)$ state, based on  $\Delta\phi^{t\bar t}_{\ell\ell}$ distribution. 
} 
\end{figure}

In Fig.~\ref{fig:tthCP_bound} (left panel) we display the  95\% C.L. bound to distinguish the CP-$\alpha$ Higgs-top interaction from the SM via $t\bar{t}h$ production. Our limits are based on a binned log-likelihood analysis invoking the CL$_s$ method for  $(\Delta\phi^{\rm lab}_{\ell\ell},\theta^*)$ (blue dashed), and $(\Delta\phi^{\rm lab}_{\ell\ell},\theta^*,\Delta\phi^{t\bar t}_{\ell\ell})$ (blue full)~\cite{Read:2002hq}. 
The bounds are obtained, including backgrounds, parton-shower, hadronization and semi-realistic detector effects. To illustrate the robustness of the top reconstruction method when going from the parton to the detector level, we also show the bounds using the parton-level distributions $(\Delta\phi^{\rm lab}_{\ell\ell},\theta^*)$  with the rates rescaled to the full detector analysis (black full).  
The red-solid curve, labelled as ``$(\Delta\phi^{\rm lab}_{\ell\ell}, \,\Delta\Phi_{jj})$", was extracted for comparison from Ref.~\cite{Buckley:2015vsa}, which runs a different analysis.
To  focus only  on  measurement  of the CP-phase,  we  fix  the  number  of  signal $t\bar{t}h$ events to the SM prediction $\alpha=0$, comparing only the shapes between the null and pseudo-hypotheses. We note that the top reconstruction in the dileptonic channel, where the top spin analyzing power is maximal, results in relevant sensitivity improvements  for the direct Higgs-top CP-phase measurement. 
While the lab-observables  $(\Delta\phi^{\rm lab}_{\ell\ell},\Delta\phi_{jj})$ result in the limit $\cos\alpha<0.5$ at 95\% CL
for the high-lumi LHC with 3~ab$^{-1}$, the addition of our observables defined at the top pair rest frame in two scenarios $(\Delta\phi^{\rm lab}_{\ell\ell},\theta^*)$ and  $(\Delta\phi^{\rm lab}_{\ell\ell},\Delta\phi^{t\bar t}_{\ell\ell},\theta^*)$, result in relevant improvements of  ${\cos\alpha<0.65}$ and   ${\cos\alpha<0.7}$, respectively. 
 
As we are able to probe $\Delta\phi^{t\bar t}_{\ell\ell}$, that is sensitive to the sign of $\alpha$, we can go beyond and inquire  if the LHC will be able to capture also the CP-phase sign.  In Fig.~\ref{fig:tthCP_bound} (right panel), we show the luminosity needed to disentangle the CP$(\alpha=\frac{\pi}{4})$ from the CP$(\alpha=-\frac{\pi}{4})$ state based on $\Delta\phi^{t\bar t}_{\ell\ell}$ distribution. We chose $\pm\frac{\pi}{4}$ for an illustration, since they give the largest difference. The observation of the sign for the maximal CP violation case requires at least 8 ab$^{-1}$ of data at the 14 TeV LHC even at 1$\sigma$-level.

\section{Summary\label{sec:summary}}

Characterizing the Higgs boson is a critical component of the LHC program. In this paper, we have studied the
\emph{direct} Higgs-top CP-phase determination via the $t\bar{t}h$ channel with Higgs decaying to bottom quarks 
and the top-quarks  in the dileptonic mode. Although this $t\bar t$ decay mode leads to maximal spin analyzing power, 
it always accompanies two neutrinos in the final state, making the analysis and reconstruction challenging.

We show that kinematic reconstruction can be obtained via the $M_2$ algorithm. This method is entirely based on mass minimization,
being more flexible for new physics studies and robust for our spin-correlation analysis. We expanded the previous $M_2$-assisted 
reconstruction studies, investigating effects of parton-shower, hadronization and detector resolution. We found that the algorithm 
performance in resolving two fold ambiguity still remains superior despite the slightly worse momentum reconstruction when compared
to the parton level. We prove however that an additional mass selection on $M_{2CW}^{(b\ell)}$ can efficiently improve the reconstruction
efficiencies.

We then studied the Higgs-top CP-phase discrimination via a realistic Monte Carlo analysis.  We show that the CP sensitivity of the 
azimuthal angle between two leptons in the laboratory frame $\Delta\phi^{\rm lab}_{\ell\ell}$ can be relevantly enhanced when combined
with $t \bar t$ rest of frame observables:   top quark production angle $\theta^\ast$ and  $\Delta \phi^{t \bar{t}}_{\ell\ell }$, where the 
latter is a truly CP-odd observable, sensitive to the sign of the CP-phase. Including the relevant backgrounds, we have performed a binned log-likelihood  analysis and computed the luminosity required to distinguish the SM Higgs from an arbitrary CP-phase at 95\% confidence level.  Based on our results, the Higgs-top CP-phase can be probed up to $\cos\alpha< 0.7$ at the high luminosity LHC.

%
\section*{Acknowledgments}
We are grateful to HTCaaS group of the Korea Institute of Science and Technology Information (KISTI) for providing the necessary 
computing resources. KK thanks the PITT-PACC for hospitality and support during the initial stage of this work. DG was funded by 
U.S. National Science Foundation under the grant PHY-1519175.  This work is supported in part by U.S. National Science Foundation 
(PHY-1519175) and U.S. Department of Energy (DE-SC0017965, DE-SC0017988).

\appendix

\section{Parameterization of detector resolution effects}
\label{app:DR}

The jet energy resolution is parametrized by a noise ($N$), a stochastic ($S$), and a constant ($C$) terms
\begin{align}
\frac{\sigma}{E} =  \sqrt{
\bigg( \frac{N}{E} \bigg)^2 +\bigg( \frac{S}{\sqrt{E}}\bigg)^2  +C^2~,
}
\end{align}
where in our analysis we use $N=5.3$, $S=0.74$ and $C=0.05$ respectively \cite{ATL-PHYS-PUB-2013-004}. 

The electron energy resolution is based on the parameterization
\begin{align}
\frac{\sigma}{E} =  \frac{0.3}{E} + \frac{0.1}{\sqrt{E}} + 0.01 ~.
\end{align}

The muon energy resolution is derived by the Inner Detector (ID) and Muon Spectrometer (MS) resolution functions
\begin{align}
\sigma = \frac{ \sigma_{\text{ID}}~ \sigma_{\text{MS} }  } { \sqrt{ \sigma^2_{\text{ID}} + \sigma^2_{\text{MS}   } } }~,
\end{align}
where
\bea
\label{eq:MuonSmear}
\sigma_{\text{ID}} &=& E~\sqrt{ a^2_1 + ( a_2 ~ E )^2   } \\
\sigma_{\text{MS}} &=& E~\sqrt{ \bigg( \frac{b_0}{E} \bigg)^2 + b^2_1 + (b_2~E)^2   }~\;.
\eea
We choose $a_1 = 0.023035$, $a_2 = 0.000347$, $b_0 = 0.12$, $b_1 = 0.03278$ and $b_2 = 0.00014$ in our study \cite{ATL-PHYS-PUB-2013-004}.

\bibliographystyle{JHEP}
\bibliography{draft}

\providecommand{\href}[2]{#2}\begingroup\raggedright\begin{thebibliography}{10}

\bibitem{Aad:2012tfa}
{\scshape ATLAS} collaboration, G.~Aad et~al., \emph{{Observation of a new
  particle in the search for the Standard Model Higgs boson with the ATLAS
  detector at the LHC}},
  \href{http://dx.doi.org/10.1016/j.physletb.2012.08.020}{\emph{Phys.Lett.}
  {\bf B716} (2012) 1--29}, [\href{https://arxiv.org/abs/1207.7214}{{\tt
  1207.7214}}].

\bibitem{Chatrchyan:2012ufa}
{\scshape CMS} collaboration, S.~Chatrchyan et~al., \emph{{Observation of a new
  boson at a mass of 125 GeV with the CMS experiment at the LHC}},
  \href{http://dx.doi.org/10.1016/j.physletb.2012.08.021}{\emph{Phys.Lett.}
  {\bf B716} (2012) 30--61}, [\href{https://arxiv.org/abs/1207.7235}{{\tt
  1207.7235}}].

\bibitem{Higgs:1964ia}
P.~W. Higgs, \emph{{Broken symmetries, massless particles and gauge fields}},
  \href{http://dx.doi.org/10.1016/0031-9163(64)91136-9}{\emph{Phys. Lett.} {\bf
  12} (1964) 132--133}.

\bibitem{Higgs:1964pj}
P.~W. Higgs, \emph{{Broken Symmetries and the Masses of Gauge Bosons}},
  \href{http://dx.doi.org/10.1103/PhysRevLett.13.508}{\emph{Phys. Rev. Lett.}
  {\bf 13} (1964) 508--509}.

\bibitem{Englert:1964et}
F.~Englert and R.~Brout, \emph{{Broken Symmetry and the Mass of Gauge Vector
  Mesons}}, \href{http://dx.doi.org/10.1103/PhysRevLett.13.321}{\emph{Phys.
  Rev. Lett.} {\bf 13} (1964) 321--323}.

\bibitem{Khachatryan:2016vau}
{\scshape ATLAS, CMS} collaboration, G.~Aad et~al., \emph{{Measurements of the
  Higgs boson production and decay rates and constraints on its couplings from
  a combined ATLAS and CMS analysis of the LHC pp collision data at $
  \sqrt{s}=7 $ and 8 TeV}},
  \href{http://dx.doi.org/10.1007/JHEP08(2016)045}{\emph{JHEP} {\bf 08} (2016)
  045}, [\href{https://arxiv.org/abs/1606.02266}{{\tt 1606.02266}}].

\bibitem{Corbett:2015ksa}
T.~Corbett, O.~J.~P. Eboli, D.~Goncalves, J.~Gonzalez-Fraile, T.~Plehn and
  M.~Rauch, \emph{{The Higgs Legacy of the LHC Run I}},
  \href{http://dx.doi.org/10.1007/JHEP08(2015)156}{\emph{JHEP} {\bf 08} (2015)
  156}, [\href{https://arxiv.org/abs/1505.05516}{{\tt 1505.05516}}].

\bibitem{Sakharov:1967dj}
A.~D. Sakharov, \emph{{Violation of CP Invariance, c Asymmetry, and Baryon
  Asymmetry of the Universe}},
  \href{http://dx.doi.org/10.1070/PU1991v034n05ABEH002497}{\emph{Pisma Zh.
  Eksp. Teor. Fiz.} {\bf 5} (1967) 32--35}.

\bibitem{Espinosa:2011eu}
J.~R. Espinosa, B.~Gripaios, T.~Konstandin and F.~Riva, \emph{{Electroweak
  Baryogenesis in Non-minimal Composite Higgs Models}},
  \href{http://dx.doi.org/10.1088/1475-7516/2012/01/012}{\emph{JCAP} {\bf 1201}
  (2012) 012}, [\href{https://arxiv.org/abs/1110.2876}{{\tt 1110.2876}}].

\bibitem{Plehn:2001nj}
T.~Plehn, D.~L. Rainwater and D.~Zeppenfeld, \emph{{Determining the structure
  of Higgs couplings at the LHC}},
  \href{http://dx.doi.org/10.1103/PhysRevLett.88.051801}{\emph{Phys. Rev.
  Lett.} {\bf 88} (2002) 051801},
  [\href{https://arxiv.org/abs/hep-ph/0105325}{{\tt hep-ph/0105325}}].

\bibitem{Hagiwara:2009wt}
K.~Hagiwara, Q.~Li and K.~Mawatari, \emph{{Jet angular correlation in
  vector-boson fusion processes at hadron colliders}},
  \href{http://dx.doi.org/10.1088/1126-6708/2009/07/101}{\emph{JHEP} {\bf 07}
  (2009) 101}, [\href{https://arxiv.org/abs/0905.4314}{{\tt 0905.4314}}].

\bibitem{Bolognesi:2012mm}
S.~Bolognesi, Y.~Gao, A.~V. Gritsan, K.~Melnikov, M.~Schulze, N.~V. Tran
  et~al., \emph{{On the spin and parity of a single-produced resonance at the
  LHC}}, \href{http://dx.doi.org/10.1103/PhysRevD.86.095031}{\emph{Phys. Rev.}
  {\bf D86} (2012) 095031}, [\href{https://arxiv.org/abs/1208.4018}{{\tt
  1208.4018}}].

\bibitem{Englert:2012xt}
C.~Englert, D.~Goncalves-Netto, K.~Mawatari and T.~Plehn, \emph{{Higgs Quantum
  Numbers in Weak Boson Fusion}},
  \href{http://dx.doi.org/10.1007/JHEP01(2013)148}{\emph{JHEP} {\bf 01} (2013)
  148}, [\href{https://arxiv.org/abs/1212.0843}{{\tt 1212.0843}}].

\bibitem{Freitas:2012kw}
A.~Freitas and P.~Schwaller, \emph{{Higgs CP Properties From Early LHC Data}},
  \href{http://dx.doi.org/10.1103/PhysRevD.87.055014}{\emph{Phys. Rev.} {\bf
  D87} (2013) 055014}, [\href{https://arxiv.org/abs/1211.1980}{{\tt
  1211.1980}}].

\bibitem{Ellis:2012wg}
J.~Ellis and D.~S. Hwang, \emph{{Does the `Higgs' have Spin Zero?}},
  \href{http://dx.doi.org/10.1007/JHEP09(2012)071}{\emph{JHEP} {\bf 09} (2012)
  071}, [\href{https://arxiv.org/abs/1202.6660}{{\tt 1202.6660}}].

\bibitem{Ellis:2012jv}
J.~Ellis, R.~Fok, D.~S. Hwang, V.~Sanz and T.~You, \emph{{Distinguishing
  'Higgs' spin hypotheses using $\gamma \gamma$ and $W W^*$ decays}},
  \href{http://dx.doi.org/10.1140/epjc/s10052-013-2488-5}{\emph{Eur. Phys. J.}
  {\bf C73} (2013) 2488}, [\href{https://arxiv.org/abs/1210.5229}{{\tt
  1210.5229}}].

\bibitem{Englert:2013opa}
C.~Englert, D.~Goncalves, G.~Nail and M.~Spannowsky, \emph{{The shape of
  spins}}, \href{http://dx.doi.org/10.1103/PhysRevD.88.013016}{\emph{Phys.
  Rev.} {\bf D88} (2013) 013016}, [\href{https://arxiv.org/abs/1304.0033}{{\tt
  1304.0033}}].

\bibitem{Khachatryan:2014kca}
{\scshape CMS} collaboration, V.~Khachatryan et~al., \emph{{Constraints on the
  spin-parity and anomalous HVV couplings of the Higgs boson in proton
  collisions at 7 and 8 TeV}},
  \href{http://dx.doi.org/10.1103/PhysRevD.92.012004}{\emph{Phys. Rev.} {\bf
  D92} (2015) 012004}, [\href{https://arxiv.org/abs/1411.3441}{{\tt
  1411.3441}}].

\bibitem{Brehmer:2017lrt}
J.~Brehmer, F.~Kling, T.~Plehn and T.~M.~P. Tait, \emph{{Better Higgs-CP Tests
  Through Information Geometry}},  \href{https://arxiv.org/abs/1712.02350}{{\tt
  1712.02350}}.

\bibitem{Buchmuller:1985jz}
W.~Buchmuller and D.~Wyler, \emph{{Effective Lagrangian Analysis of New
  Interactions and Flavor Conservation}},
  \href{http://dx.doi.org/10.1016/0550-3213(86)90262-2}{\emph{Nucl. Phys.} {\bf
  B268} (1986) 621--653}.

\bibitem{Grzadkowski:2010es}
B.~Grzadkowski, M.~Iskrzynski, M.~Misiak and J.~Rosiek, \emph{{Dimension-Six
  Terms in the Standard Model Lagrangian}},
  \href{http://dx.doi.org/10.1007/JHEP10(2010)085}{\emph{JHEP} {\bf 10} (2010)
  085}, [\href{https://arxiv.org/abs/1008.4884}{{\tt 1008.4884}}].

\bibitem{Ellis:2013yxa}
J.~Ellis, D.~S. Hwang, K.~Sakurai and M.~Takeuchi, \emph{{Disentangling
  Higgs-Top Couplings in Associated Production}},
  \href{http://dx.doi.org/10.1007/JHEP04(2014)004}{\emph{JHEP} {\bf 04} (2014)
  004}, [\href{https://arxiv.org/abs/1312.5736}{{\tt 1312.5736}}].

\bibitem{Buckley:2015vsa}
M.~R. Buckley and D.~Goncalves, \emph{{Boosting the Direct CP Measurement of
  the Higgs-Top Coupling}},
  \href{http://dx.doi.org/10.1103/PhysRevLett.116.091801}{\emph{Phys. Rev.
  Lett.} {\bf 116} (2016) 091801},
  [\href{https://arxiv.org/abs/1507.07926}{{\tt 1507.07926}}].

\bibitem{Boudjema:2015nda}
F.~Boudjema, R.~M. Godbole, D.~Guadagnoli and K.~A. Mohan, \emph{{Lab-frame
  observables for probing the top-Higgs interaction}},
  \href{http://dx.doi.org/10.1103/PhysRevD.92.015019}{\emph{Phys. Rev.} {\bf
  D92} (2015) 015019}, [\href{https://arxiv.org/abs/1501.03157}{{\tt
  1501.03157}}].

\bibitem{Mileo:2016mxg}
N.~Mileo, K.~Kiers, A.~Szynkman, D.~Crane and E.~Gegner, \emph{{Pseudoscalar
  top-Higgs coupling: exploration of CP-odd observables to resolve the sign
  ambiguity}}, \href{http://dx.doi.org/10.1007/JHEP07(2016)056}{\emph{JHEP}
  {\bf 07} (2016) 056}, [\href{https://arxiv.org/abs/1603.03632}{{\tt
  1603.03632}}].

\bibitem{Gritsan:2016hjl}
A.~V. Gritsan, R.~Röntsch, M.~Schulze and M.~Xiao, \emph{{Constraining
  anomalous Higgs boson couplings to the heavy flavor fermions using matrix
  element techniques}},
  \href{http://dx.doi.org/10.1103/PhysRevD.94.055023}{\emph{Phys. Rev.} {\bf
  D94} (2016) 055023}, [\href{https://arxiv.org/abs/1606.03107}{{\tt
  1606.03107}}].

\bibitem{Berge:2008wi}
S.~Berge, W.~Bernreuther and J.~Ziethe, \emph{{Determining the CP parity of
  Higgs bosons at the LHC in their tau decay channels}},
  \href{http://dx.doi.org/10.1103/PhysRevLett.100.171605}{\emph{Phys. Rev.
  Lett.} {\bf 100} (2008) 171605}, [\href{https://arxiv.org/abs/0801.2297}{{\tt
  0801.2297}}].

\bibitem{Harnik:2013aja}
R.~Harnik, A.~Martin, T.~Okui, R.~Primulando and F.~Yu, \emph{{Measuring CP
  violation in $h \to \tau^+ \tau^-$ at colliders}},
  \href{http://dx.doi.org/10.1103/PhysRevD.88.076009}{\emph{Phys. Rev.} {\bf
  D88} (2013) 076009}, [\href{https://arxiv.org/abs/1308.1094}{{\tt
  1308.1094}}].

\bibitem{Dolan:2016qvg}
M.~J. Dolan, M.~Spannowsky, Q.~Wang and Z.-H. Yu, \emph{{Determining the
  quantum numbers of simplified models in $t\bar{t}X$ production at the LHC}},
  \href{http://dx.doi.org/10.1103/PhysRevD.94.015025}{\emph{Phys. Rev.} {\bf
  D94} (2016) 015025}, [\href{https://arxiv.org/abs/1606.00019}{{\tt
  1606.00019}}].

\bibitem{Santos:2015dja}
S.~P. Amor~dos Santos et~al., \emph{{Angular distributions in $t
  \overline{t}H(H ? b \overline{b})$ reconstructed events at the LHC}},
  \href{http://dx.doi.org/10.1103/PhysRevD.92.034021}{\emph{Phys. Rev.} {\bf
  D92} (2015) 034021}, [\href{https://arxiv.org/abs/1503.07787}{{\tt
  1503.07787}}].

\bibitem{Goncalves:2016qhh}
D.~Goncalves and D.~Lopez-Val, \emph{{Pseudoscalar searches with dileptonic
  tops and jet substructure}},
  \href{http://dx.doi.org/10.1103/PhysRevD.94.095005}{\emph{Phys. Rev.} {\bf
  D94} (2016) 095005}, [\href{https://arxiv.org/abs/1607.08614}{{\tt
  1607.08614}}].

\bibitem{Han:2016bvf}
T.~Han, S.~Mukhopadhyay, B.~Mukhopadhyaya and Y.~Wu, \emph{{Measuring the CP
  property of Higgs coupling to tau leptons in the VBF channel at the LHC}},
  \href{http://dx.doi.org/10.1007/JHEP05(2017)128}{\emph{JHEP} {\bf 05} (2017)
  128}, [\href{https://arxiv.org/abs/1612.00413}{{\tt 1612.00413}}].

\bibitem{Hagiwara:2016zqz}
K.~Hagiwara, K.~Ma and S.~Mori, \emph{{Probing CP violation in $h\to
  \tau^{-}\tau^{+}$ at the LHC}},
  \href{http://dx.doi.org/10.1103/PhysRevLett.118.171802}{\emph{Phys. Rev.
  Lett.} {\bf 118} (2017) 171802},
  [\href{https://arxiv.org/abs/1609.00943}{{\tt 1609.00943}}].

\bibitem{Brod:2013cka}
J.~Brod, U.~Haisch and J.~Zupan, \emph{{Constraints on CP-violating Higgs
  couplings to the third generation}},
  \href{http://dx.doi.org/10.1007/JHEP11(2013)180}{\emph{JHEP} {\bf 11} (2013)
  180}, [\href{https://arxiv.org/abs/1310.1385}{{\tt 1310.1385}}].

\bibitem{DelDuca:2006hk}
V.~Del~Duca, G.~Klamke, D.~Zeppenfeld, M.~L. Mangano, M.~Moretti, F.~Piccinini
  et~al., \emph{{Monte Carlo studies of the jet activity in Higgs + 2 jet
  events}}, \href{http://dx.doi.org/10.1088/1126-6708/2006/10/016}{\emph{JHEP}
  {\bf 10} (2006) 016}, [\href{https://arxiv.org/abs/hep-ph/0608158}{{\tt
  hep-ph/0608158}}].

\bibitem{Dolan:2014upa}
M.~J. Dolan, P.~Harris, M.~Jankowiak and M.~Spannowsky, \emph{{Constraining
  $CP$-violating Higgs Sectors at the LHC using gluon fusion}},
  \href{http://dx.doi.org/10.1103/PhysRevD.90.073008}{\emph{Phys. Rev.} {\bf
  D90} (2014) 073008}, [\href{https://arxiv.org/abs/1406.3322}{{\tt
  1406.3322}}].

\bibitem{Banfi:2013yoa}
A.~Banfi, A.~Martin and V.~Sanz, \emph{{Probing top-partners in Higgs+jets}},
  \href{http://dx.doi.org/10.1007/JHEP08(2014)053}{\emph{JHEP} {\bf 08} (2014)
  053}, [\href{https://arxiv.org/abs/1308.4771}{{\tt 1308.4771}}].

\bibitem{Azatov:2013xha}
A.~Azatov and A.~Paul, \emph{{Probing Higgs couplings with high $p_T$ Higgs
  production}}, \href{http://dx.doi.org/10.1007/JHEP01(2014)014}{\emph{JHEP}
  {\bf 01} (2014) 014}, [\href{https://arxiv.org/abs/1309.5273}{{\tt
  1309.5273}}].

\bibitem{Grojean:2013nya}
C.~Grojean, E.~Salvioni, M.~Schlaffer and A.~Weiler, \emph{{Very boosted Higgs
  in gluon fusion}},
  \href{http://dx.doi.org/10.1007/JHEP05(2014)022}{\emph{JHEP} {\bf 05} (2014)
  022}, [\href{https://arxiv.org/abs/1312.3317}{{\tt 1312.3317}}].

\bibitem{Schlaffer:2014osa}
M.~Schlaffer, M.~Spannowsky, M.~Takeuchi, A.~Weiler and C.~Wymant,
  \emph{{Boosted Higgs Shapes}},
  \href{http://dx.doi.org/10.1140/epjc/s10052-014-3120-z}{\emph{Eur. Phys. J.}
  {\bf C74} (2014) 3120}, [\href{https://arxiv.org/abs/1405.4295}{{\tt
  1405.4295}}].

\bibitem{Buschmann:2014twa}
M.~Buschmann, C.~Englert, D.~Goncalves, T.~Plehn and M.~Spannowsky,
  \emph{{Resolving the Higgs-Gluon Coupling with Jets}},
  \href{http://dx.doi.org/10.1103/PhysRevD.90.013010}{\emph{Phys. Rev.} {\bf
  D90} (2014) 013010}, [\href{https://arxiv.org/abs/1405.7651}{{\tt
  1405.7651}}].

\bibitem{Buschmann:2014sia}
M.~Buschmann, D.~Goncalves, S.~Kuttimalai, M.~Schonherr, F.~Krauss and
  T.~Plehn, \emph{{Mass Effects in the Higgs-Gluon Coupling: Boosted vs
  Off-Shell Production}},
  \href{http://dx.doi.org/10.1007/JHEP02(2015)038}{\emph{JHEP} {\bf 02} (2015)
  038}, [\href{https://arxiv.org/abs/1410.5806}{{\tt 1410.5806}}].

\bibitem{ATLAS-CONF-2017-077}
{\scshape ATLAS} collaboration, \emph{{Evidence for the associated production
  of the Higgs boson and a top quark pair with the ATLAS detector}},  Tech.
  Rep. ATLAS-CONF-2017-077, CERN, Geneva, Nov, 2017.

\bibitem{Sirunyan:2018hoz}
{\scshape CMS} collaboration, A.~M. Sirunyan et~al., \emph{{Observation of
  $\mathrm{t\overline{t}}$H production}},
  \href{https://arxiv.org/abs/1804.02610}{{\tt 1804.02610}}.

\bibitem{CMS:2013xfa}
{\scshape CMS} collaboration, \emph{{Projected Performance of an Upgraded CMS
  Detector at the LHC and HL-LHC: Contribution to the Snowmass Process}},  in
  \emph{{Proceedings, 2013 Community Summer Study on the Future of U.S.
  Particle Physics: Snowmass on the Mississippi (CSS2013): Minneapolis, MN,
  USA, July 29-August 6, 2013}}, 2013.
\newblock \href{https://arxiv.org/abs/1307.7135}{{\tt 1307.7135}}.

\bibitem{Debnath:2017ktz}
D.~Debnath, D.~Kim, J.~H. Kim, K.~Kong and K.~T. Matchev, \emph{{Resolving
  Combinatorial Ambiguities in Dilepton $t\bar t$ Event Topologies with
  Constrained $M_2$ Variables}},
  \href{http://dx.doi.org/10.1103/PhysRevD.96.076005}{\emph{Phys. Rev.} {\bf
  D96} (2017) 076005}, [\href{https://arxiv.org/abs/1706.04995}{{\tt
  1706.04995}}].

\bibitem{Mahlon:1995zn}
G.~Mahlon and S.~J. Parke, \emph{{Angular correlations in top quark pair
  production and decay at hadron colliders}},
  \href{http://dx.doi.org/10.1103/PhysRevD.53.4886}{\emph{Phys. Rev.} {\bf D53}
  (1996) 4886--4896}, [\href{https://arxiv.org/abs/hep-ph/9512264}{{\tt
  hep-ph/9512264}}].

\bibitem{ATLAS:2016jct}
{\scshape ATLAS} collaboration, \emph{{Measurements of $t\bar{t}$ differential
  cross-sections in the all-hadronic channel with the ATLAS detector using
  highly boosted top quarks in $pp$ collisions at $\sqrt{s}=13$ TeV}},
  {\emph{ATLAS-CONF-2016-100} (2016) }.

\bibitem{Lester:1999tx}
C.~G. Lester and D.~J. Summers, \emph{{Measuring masses of semiinvisibly
  decaying particles pair produced at hadron colliders}},
  \href{http://dx.doi.org/10.1016/S0370-2693(99)00945-4}{\emph{Phys. Lett.}
  {\bf B463} (1999) 99--103}, [\href{https://arxiv.org/abs/hep-ph/9906349}{{\tt
  hep-ph/9906349}}].

\bibitem{Konar:2009wn}
P.~Konar, K.~Kong, K.~T. Matchev and M.~Park, \emph{{Superpartner Mass
  Measurement Technique using 1D Orthogonal Decompositions of the Cambridge
  Transverse Mass Variable $M_{T2}$}},
  \href{http://dx.doi.org/10.1103/PhysRevLett.105.051802}{\emph{Phys. Rev.
  Lett.} {\bf 105} (2010) 051802}, [\href{https://arxiv.org/abs/0910.3679}{{\tt
  0910.3679}}].

\bibitem{Burns:2008va}
M.~Burns, K.~Kong, K.~T. Matchev and M.~Park, \emph{{Using Subsystem MT2 for
  Complete Mass Determinations in Decay Chains with Missing Energy at Hadron
  Colliders}},
  \href{http://dx.doi.org/10.1088/1126-6708/2009/03/143}{\emph{JHEP} {\bf 03}
  (2009) 143}, [\href{https://arxiv.org/abs/0810.5576}{{\tt 0810.5576}}].

\bibitem{Konar:2009qr}
P.~Konar, K.~Kong, K.~T. Matchev and M.~Park, \emph{{Dark Matter Particle
  Spectroscopy at the LHC: Generalizing M(T2) to Asymmetric Event Topologies}},
  \href{http://dx.doi.org/10.1007/JHEP04(2010)086}{\emph{JHEP} {\bf 04} (2010)
  086}, [\href{https://arxiv.org/abs/0911.4126}{{\tt 0911.4126}}].

\bibitem{Barr:2011xt}
A.~J. Barr, T.~J. Khoo, P.~Konar, K.~Kong, C.~G. Lester, K.~T. Matchev et~al.,
  \emph{{Guide to transverse projections and mass-constraining variables}},
  \href{http://dx.doi.org/10.1103/PhysRevD.84.095031}{\emph{Phys. Rev.} {\bf
  D84} (2011) 095031}, [\href{https://arxiv.org/abs/1105.2977}{{\tt
  1105.2977}}].

\bibitem{Cho:2014naa}
W.~S. Cho, J.~S. Gainer, D.~Kim, K.~T. Matchev, F.~Moortgat, L.~Pape et~al.,
  \emph{{On-shell constrained $M_2$ variables with applications to mass
  measurements and topology disambiguation}},
  \href{http://dx.doi.org/10.1007/JHEP08(2014)070}{\emph{JHEP} {\bf 08} (2014)
  070}, [\href{https://arxiv.org/abs/1401.1449}{{\tt 1401.1449}}].

\bibitem{Cho:2015laa}
W.~S. Cho, J.~S. Gainer, D.~Kim, S.~H. Lim, K.~T. Matchev, F.~Moortgat et~al.,
  \emph{{OPTIMASS: A Package for the Minimization of Kinematic Mass Functions
  with Constraints}},
  \href{http://dx.doi.org/10.1007/JHEP01(2016)026}{\emph{JHEP} {\bf 01} (2016)
  026}, [\href{https://arxiv.org/abs/1508.00589}{{\tt 1508.00589}}].

\bibitem{Konar:2008ei}
P.~Konar, K.~Kong and K.~T. Matchev, \emph{{$\sqrt{\hat{s}}_{min}$ : A Global
  inclusive variable for determining the mass scale of new physics in events
  with missing energy at hadron colliders}},
  \href{http://dx.doi.org/10.1088/1126-6708/2009/03/085}{\emph{JHEP} {\bf 03}
  (2009) 085}, [\href{https://arxiv.org/abs/0812.1042}{{\tt 0812.1042}}].

\bibitem{Konar:2010ma}
P.~Konar, K.~Kong, K.~T. Matchev and M.~Park, \emph{{RECO level
  $\sqrt{s}_{min}$ and subsystem $\sqrt{s}_{min}$: Improved global inclusive
  variables for measuring the new physics mass scale in $\met$ events at hadron
  colliders}}, \href{http://dx.doi.org/10.1007/JHEP06(2011)041}{\emph{JHEP}
  {\bf 06} (2011) 041}, [\href{https://arxiv.org/abs/1006.0653}{{\tt
  1006.0653}}].

\bibitem{Barr:2010zj}
A.~J. Barr and C.~G. Lester, \emph{{A Review of the Mass Measurement Techniques
  proposed for the Large Hadron Collider}},
  \href{http://dx.doi.org/10.1088/0954-3899/37/12/123001}{\emph{J. Phys.} {\bf
  G37} (2010) 123001}, [\href{https://arxiv.org/abs/1004.2732}{{\tt
  1004.2732}}].

\bibitem{Kim:2017awi}
D.~Kim, K.~T. Matchev, F.~Moortgat and L.~Pape, \emph{{Testing Invisible
  Momentum Ansatze in Missing Energy Events at the LHC}},
  \href{http://dx.doi.org/10.1007/JHEP08(2017)102}{\emph{JHEP} {\bf 08} (2017)
  102}, [\href{https://arxiv.org/abs/1703.06887}{{\tt 1703.06887}}].

\bibitem{Ross:2007rm}
G.~G. Ross and M.~Serna, \emph{{Mass determination of new states at hadron
  colliders}},
  \href{http://dx.doi.org/10.1016/j.physletb.2008.06.003}{\emph{Phys. Lett.}
  {\bf B665} (2008) 212--218}, [\href{https://arxiv.org/abs/0712.0943}{{\tt
  0712.0943}}].

\bibitem{Baringer:2011nh}
P.~Baringer, K.~Kong, M.~McCaskey and D.~Noonan, \emph{{Revisiting
  Combinatorial Ambiguities at Hadron Colliders with $M_{T2}$}},
  \href{http://dx.doi.org/10.1007/JHEP10(2011)101}{\emph{JHEP} {\bf 10} (2011)
  101}, [\href{https://arxiv.org/abs/1109.1563}{{\tt 1109.1563}}].

\bibitem{Alwall:2014hca}
J.~Alwall, R.~Frederix, S.~Frixione, V.~Hirschi, F.~Maltoni, O.~Mattelaer
  et~al., \emph{{The automated computation of tree-level and next-to-leading
  order differential cross sections, and their matching to parton shower
  simulations}}, \href{http://dx.doi.org/10.1007/JHEP07(2014)079}{\emph{JHEP}
  {\bf 07} (2014) 079}, [\href{https://arxiv.org/abs/1405.0301}{{\tt
  1405.0301}}].

\bibitem{Alloul:2013bka}
A.~Alloul, N.~D. Christensen, C.~Degrande, C.~Duhr and B.~Fuks,
  \emph{{FeynRules 2.0 - A complete toolbox for tree-level phenomenology}},
  \href{http://dx.doi.org/10.1016/j.cpc.2014.04.012}{\emph{Comput. Phys.
  Commun.} {\bf 185} (2014) 2250--2300},
  [\href{https://arxiv.org/abs/1310.1921}{{\tt 1310.1921}}].

\bibitem{Ball:2013hta}
{\scshape NNPDF} collaboration, R.~D. Ball, V.~Bertone, S.~Carrazza,
  L.~Del~Debbio, S.~Forte, A.~Guffanti et~al., \emph{{Parton distributions with
  QED corrections}},
  \href{http://dx.doi.org/10.1016/j.nuclphysb.2013.10.010}{\emph{Nucl. Phys.}
  {\bf B877} (2013) 290--320}, [\href{https://arxiv.org/abs/1308.0598}{{\tt
  1308.0598}}].

\bibitem{Dawson:2002tg}
S.~Dawson, L.~H. Orr, L.~Reina and D.~Wackeroth, \emph{{Associated top quark
  Higgs boson production at the LHC}},
  \href{http://dx.doi.org/10.1103/PhysRevD.67.071503}{\emph{Phys. Rev.} {\bf
  D67} (2003) 071503}, [\href{https://arxiv.org/abs/hep-ph/0211438}{{\tt
  hep-ph/0211438}}].

\bibitem{Bredenstein:2009aj}
A.~Bredenstein, A.~Denner, S.~Dittmaier and S.~Pozzorini, \emph{{NLO QCD
  corrections to pp ---> t anti-t b anti-b + X at the LHC}},
  \href{http://dx.doi.org/10.1103/PhysRevLett.103.012002}{\emph{Phys. Rev.
  Lett.} {\bf 103} (2009) 012002}, [\href{https://arxiv.org/abs/0905.0110}{{\tt
  0905.0110}}].

\bibitem{Maltoni:2015ena}
F.~Maltoni, D.~Pagani and I.~Tsinikos, \emph{{Associated production of a
  top-quark pair with vector bosons at NLO in QCD: impact on $
  \mathrm{t}\overline{\mathrm{t}}\mathrm{H} $ searches at the LHC}},
  \href{http://dx.doi.org/10.1007/JHEP02(2016)113}{\emph{JHEP} {\bf 02} (2016)
  113}, [\href{https://arxiv.org/abs/1507.05640}{{\tt 1507.05640}}].

\bibitem{Sjostrand:2006za}
T.~Sjostrand, S.~Mrenna and P.~Z. Skands, \emph{{PYTHIA 6.4 Physics and
  Manual}}, \href{http://dx.doi.org/10.1088/1126-6708/2006/05/026}{\emph{JHEP}
  {\bf 05} (2006) 026}, [\href{https://arxiv.org/abs/hep-ph/0603175}{{\tt
  hep-ph/0603175}}].

\bibitem{Cacciari:2011ma}
M.~Cacciari, G.~P. Salam and G.~Soyez, \emph{{FastJet User Manual}},
  \href{http://dx.doi.org/10.1140/epjc/s10052-012-1896-2}{\emph{Eur. Phys. J.}
  {\bf C72} (2012) 1896}, [\href{https://arxiv.org/abs/1111.6097}{{\tt
  1111.6097}}].

\bibitem{Cacciari:2008gp}
M.~Cacciari, G.~P. Salam and G.~Soyez, \emph{{The Anti-k(t) jet clustering
  algorithm}},
  \href{http://dx.doi.org/10.1088/1126-6708/2008/04/063}{\emph{JHEP} {\bf 04}
  (2008) 063}, [\href{https://arxiv.org/abs/0802.1189}{{\tt 0802.1189}}].

\bibitem{ATL-PHYS-PUB-2013-004}
\emph{{Performance assumptions for an upgraded ATLAS detector at a
  High-Luminosity LHC}},  Tech. Rep. ATL-PHYS-PUB-2013-004, CERN, Geneva, Mar,
  2013.

\bibitem{Backovic:2012jk}
M.~Backović and J.~Juknevich, \emph{{TemplateTagger v1.0.0: A Template
  Matching Tool for Jet Substructure}},
  \href{http://dx.doi.org/10.1016/j.cpc.2013.12.018}{\emph{Comput. Phys.
  Commun.} {\bf 185} (2014) 1322--1338},
  [\href{https://arxiv.org/abs/1212.2978}{{\tt 1212.2978}}].

\bibitem{Almeida:2010pa}
L.~G. Almeida, S.~J. Lee, G.~Perez, G.~Sterman and I.~Sung, \emph{{Template
  Overlap Method for Massive Jets}},
  \href{http://dx.doi.org/10.1103/PhysRevD.82.054034}{\emph{Phys. Rev.} {\bf
  D82} (2010) 054034}, [\href{https://arxiv.org/abs/1006.2035}{{\tt
  1006.2035}}].

\bibitem{Backovic:2013bga}
M.~Backovic, O.~Gabizon, J.~Juknevich, G.~Perez and Y.~Soreq, \emph{{Measuring
  boosted tops in semi-leptonic $t\bar t$ events for the standard model and
  beyond}}, \href{http://dx.doi.org/10.1007/JHEP04(2014)176}{\emph{JHEP} {\bf
  04} (2014) 176}, [\href{https://arxiv.org/abs/1311.2962}{{\tt 1311.2962}}].

\bibitem{ATL-PHYS-PUB-2016-026}
{\scshape ATLAS Collaboration} collaboration, \emph{{Expected performance for
  an upgraded ATLAS detector at High-Luminosity LHC}},  Tech. Rep.
  ATL-PHYS-PUB-2016-026, CERN, Geneva, Oct, 2016.

\bibitem{Backovic:2015bca}
M.~Backovic, T.~Flacke, J.~H. Kim and S.~J. Lee, \emph{{Search Strategies for
  TeV Scale Fermionic Top Partners with Charge 2/3}},
  \href{http://dx.doi.org/10.1007/JHEP04(2016)014}{\emph{JHEP} {\bf 04} (2016)
  014}, [\href{https://arxiv.org/abs/1507.06568}{{\tt 1507.06568}}].

\bibitem{AmorDosSantos:2017ayi}
S.~Amor Dos~Santos et~al., \emph{{Probing the CP nature of the Higgs coupling
  in $t{\bar t}h$ events at the LHC}},
  \href{http://dx.doi.org/10.1103/PhysRevD.96.013004}{\emph{Phys. Rev.} {\bf
  D96} (2017) 013004}, [\href{https://arxiv.org/abs/1704.03565}{{\tt
  1704.03565}}].

\bibitem{Cowan:2010js}
G.~Cowan, K.~Cranmer, E.~Gross and O.~Vitells, \emph{{Asymptotic formulae for
  likelihood-based tests of new physics}},
  \href{http://dx.doi.org/10.1140/epjc/s10052-011-1554-0,
  10.1140/epjc/s10052-013-2501-z}{\emph{Eur. Phys. J.} {\bf C71} (2011) 1554},
  [\href{https://arxiv.org/abs/1007.1727}{{\tt 1007.1727}}].

\bibitem{Read:2002hq}
A.~L. Read, \emph{{Presentation of search results: The CL(s) technique}},
  \href{http://dx.doi.org/10.1088/0954-3899/28/10/313}{\emph{J. Phys.} {\bf
  G28} (2002) 2693--2704}.

\end{thebibliography}\endgroup

\end{document}